\documentclass[12pt]{article}
\usepackage[english]{babel}
\usepackage[cp1251]{inputenc}
\usepackage{doc}
\usepackage{amsmath}
\usepackage{amsfonts}   
\usepackage{cite}    
\makeatother
\numberwithin{equation}{section}
\usepackage{makeidx}
\makeindex
\makeatletter
\renewcommand{\@biblabel}[1]{#1.}

\newtheorem{thm}{Theorem}    [section]
\newtheorem{remark}{Remark}   [section]
\newtheorem{lemma}{Lemma}   [section]
\newtheorem{cor}{Corollary}   [section]

\newtheorem{defin}{Definition}   [section]

\begin{document}

\centerline{ \bf\Large{Generalization of Doob Decomposition Theorem }}
 \centerline{ \bf\Large{and  Risk Assessment in Incomplete Markets}}
\vskip 5mm
{\bf \centerline {\Large  N.S. Gonchar } }     
\vskip 5mm
\centerline{\bf {Bogolyubov Institute for Theoretical Physics of NAS of Ukraine.}}
\vskip 2mm

\begin{abstract}
In the paper, we introduce the notion of a local  regular supermartingale relative to a convex set of equivalent measures and prove for it the necessary and sufficient conditions of   optional  Doob  decomposition in the discrete case. This Theorem is a generalization of the  famous  Doob  decomposition onto the case of supermartingales relative to a convex set of equivalent measures. 
The description of all local regular supermartingales relative to a convex set of equivalent  measures  is presented. A notion of complete set of equivalent  measures is introduced. We prove that every non negative bounded supermartingale relative to a complete set of equivalent  measures  is local regular. 
A new definition of fair price of contingent claim in incomplete market   is given and a formula for fair price of Standard option of European type is found.
\end{abstract}

\centerline{{\bf Keywords:} random process, convex set of equivalent measures, optional} 
\centerline{ Doob decomposition, regular supermartingale, martingale, }
\centerline{fair price of contingent claim}
\centerline{{\bf  2010 MSC}  60G07, 60G42.}

\section{Introduction.}

In the paper, martingales and supermartingales relative to a convex set of equivalent measures are systematically studied.  The notion of local regular supermartingale relative to a convex set of equivalent measures is introduced and the necessary and sufficient are found under that a supermartingale is local regular one. Complete description of   local regular supermartingales is given. The notion of complete convex set of equivalent measures is introduced and  established that every nonnegative supermartingale is local regular relative this set of measures.  The notion of local regular supermartingale is used for definition of fair price of contingent claim relative to a convex set of equivalent measures. Sufficient conditions of the  existence of fair price of contingent claim relative to a convex set of equivalent measures are presented. All these notions are used in the case as a convex set of equivalent measures is a set of equivalent martingale measures for evolution as risk and non risk assets.  Formulas for fair price of standard contract with option of European type  in incomplete are found. 

 The notion of complete convex set of equivalent measures permits  to give a new proof of  optional decomposition for non negative supermartingale. This proof do not use no-arbitrage arguments and measurable choice \cite{Kramkov}, \cite{FolmerKramkov1}, \cite{FolmerKabanov1},  \cite{FolmerKabanov}.

First, optional decomposition for supermartingale was opened by  by  El Karoui N. and  Quenez M. C. \cite{KarouiQuenez} for diffusion processes. After that, Kramkov D. O. and Follmer H. \cite{Kramkov}, \cite{FolmerKramkov1} proved the optional decomposition for nonnegative bounded supermartingales.  Folmer H. and Kabanov Yu. M.  \cite{FolmerKabanov1},  \cite{FolmerKabanov}  proved analogous result for an arbitrary supermartingale. Recently, Bouchard B. and Nutz M. \cite{Bouchard1} considered a class of discrete models and proved the necessary and sufficient conditions for validity of optional decomposition. 

The optional decomposition for supermartingales plays fundamental role for risk assessment in incomplete markets  \cite{KarouiQuenez}, \cite{Kramkov}, \cite{FolmerKramkov1}, \cite{Gonchar2},     
 \cite{Gonchar555},  \cite{Gonchar557}. 
Considered in the paper problem is generalization of corresponding one  that  appeared in mathematical finance  about optional decomposition for supermartingale and which is related with construction of superhedge strategy in incomplete financial markets.

Our statement of the problem unlike the above-mentioned one and it  is more general:  a supermartingale relative to a convex set of equivalent  measures is given  and it is necessary to find conditions on the supermartingale and the set of measures  under that  optional decomposition exists.

Generality of our statement of the problem is that we do not require that the considered  set of measures was generated by random process that is a local martingale  as it is done in the papers \cite{Bouchard1, KarouiQuenez, Kramkov, FolmerKabanov} and that is important for the proof of the  optional decomposition in these papers.

\section{Optional decomposition for supermartingales relative to  a convex set of equivalent measures.}

We assume that on a measurable space $\{\Omega,\mathcal{F}\}$ a filtration    ${\mathcal{F}_{m}\subset\mathcal{F}_{m+1}}\subset\mathcal{F}, \ m=\overline{0, \infty},$ and a family of measures $ M$ on $\mathcal{F}$ are given. Further, we assume that ${\cal F}_0=\{\emptyset, \Omega \}.$
A random process $\psi={\{\psi_{m}\}_{m=0}^{\infty}}$ is said to be adapted one relative to the filtration $\{{\cal F}_m\}_{m=0}^{\infty}$ if $\psi_{m}$ is ${\cal F}_m$ measurable random value for all $m=\overline{0,\infty}.$
\begin{defin}
An adapted random process  $f={\{f_{m}\}_{m=0}^{\infty}}$ is said to be   a supermartingale relative to the filtration ${\cal F}_m,\ m=\overline{0,\infty},$ and the  family of measures   $ M$  if $E^P|f_m|<\infty, \ m=\overline{1, \infty}, \ P \in M,$ and the inequalities 
\begin{eqnarray}\label{pk11} 
E^P\{f_m|{\cal F}_k\} \leq f_k, \quad 0 \leq k \leq m, \quad m=\overline{1, \infty}, \quad P \in M,
\end{eqnarray}
are valid.
\end{defin}
We consider that the filtration  ${\cal F}_m,\ m=\overline{0,\infty},$ is fixed. Further,   for a supermartingale  $f$ we use as denotation $\{f_{m}, {\cal F}_m\}_{m=0}^{\infty} $ and denotation  $\{f_{m}\}_{m=0}^{\infty}.$

Below, in a few theorems, we consider a  convex set of equivalent measures $M$ satisfying conditions: Radon -- Nicodym derivative of any measure $Q_1 \in M$ with respect to any measure  $Q_2 \in M$ satisfies inequalities 
\begin{eqnarray}\label{gon1}
0< {l\leq\frac{dQ_{1}}{dQ_{2}}\leq L}< \infty,\quad  Q_1, \   Q_2 \in M,  
\end{eqnarray}
where real numbers    $l,\ L$ do not depend on $Q_1, \ Q_2 \in M.$

 \begin{thm}\label{t1} Let ${\{f_{m}, {\cal F}_m\}_{m=0}^{\infty} }$ be a supermartingale concerning a convex set of equivalent measures $M$ satisfying conditions (\ref{gon1}).  If  for a certain measure  $P_{1}\in M$ there exist a natural number $1 \leq m_0<\infty,$  and ${\cal F}_{m_0-1}$ measurable nonnegative random value $  \varphi_{m_0},$  $P_1(\varphi_{m_0}>0)>0,$  such that   the inequality
$$f_{m_0-1}-E^{P_1}\{f_{m_0}|\mathcal{F}_{m_0-1}\}\geq \varphi_{m_0},$$
is valid, then  
\begin{eqnarray*}
f_{m_0-1}-E^{Q}\{f_{m_0}|\mathcal{F}_{m_0-1}\}\geq \frac{l }{1+L}\varphi_{m_0}, \quad   Q\in M_{\bar \varepsilon_0},
\end{eqnarray*}
where
$$  M_{\bar \varepsilon_0}=\{Q\in M, \ Q=(1-\alpha)P_{1}+\alpha P_{2},\ 0\leq\alpha \leq  \bar \varepsilon_0, \ P_{2}\in M \},  \quad P_{1}\in M,$$
 $$\bar \varepsilon_0=\frac{L}{1+L}. $$
\end{thm}

{\bf Proof.} Let   $B\in\mathcal{F}_{m_0-1}$ and  $Q=(1-\alpha)P_{1}+\alpha P_{2}, \ P_2 \in M,  \ 0<\alpha<1.$
Then 
\begin{eqnarray*}
\int\limits_{B}[f_{m_0-1}-E^{Q}\{f_{m_0}|\mathcal{F}_{m_0-1}\}]dQ=
\end{eqnarray*}
\begin{eqnarray*}
\int\limits_{B}E^{Q}\{[f_{m_0-1}-f_{m_0}]|\mathcal{F}_{m_0-1}\}dQ=
\end{eqnarray*}
\begin{eqnarray*}
\int\limits_{B}[f_{m_0-1}-f_{m_0}]dQ=
\end{eqnarray*}
\begin{eqnarray*}
(1-\alpha)\int\limits_{B}[f_{m_0-1}-f_{m_0}]dP_1+
\end{eqnarray*}
\begin{eqnarray*}
\alpha\int\limits_{B}[f_{m_0-1}-f_{m_0}]dP_2=
\end{eqnarray*}
\begin{eqnarray*} (1- \alpha)\int\limits_{B}[f_{m_0-1}-E^{P_1}\{f_{m_0}|\mathcal{F}_{m_0-1}\}]dP_1+
\end{eqnarray*}
\begin{eqnarray*}
\alpha\int\limits_{B}[f_{m_0-1}-E^{P_2}\{f_{m_0}|\mathcal{F}_{m_0-1}\}]dP_2\geq
\end{eqnarray*}
\begin{eqnarray*}
 (1- \alpha)\int\limits_{B}[f_{m_0-1}-E^{P_1}\{f_{m_0}|\mathcal{F}_{m_0-1}\}]dP_1=
\end{eqnarray*}
\begin{eqnarray*} (1- \alpha)\int\limits_{B}[f_{m_0-1}-E^{P_1}\{f_{m_0}|\mathcal{F}_{m_0-1}\}]\frac{dP_1}{dQ} dQ \geq
\end{eqnarray*}
$$  (1- \alpha) l \int\limits_{B} \varphi_{m_0}dQ\geq (1- \bar \varepsilon_0) l \int\limits_{B} \varphi_{m_0}dQ=\frac{l}{1+L}\int\limits_{B} \varphi_{m_0}dQ.$$
Arbitrariness of $B \in {\cal F}_{m_0-1}$ proves the needed inequality.

\begin{lemma}\label{l1} Any supermartingale  ${\{f_m, {\cal F}_m\}_{m=0}^{\infty}}$ relative to  a family of measures  $ M$ for which there hold equalities   $E^{P}f_{m}=f_{0}, \ m=\overline{1,\infty},$ \  ${ P\in M},$ is a martingale  with respect to this family of measures and the filtration   ${\cal F}_m,\ m=\overline{1,\infty}.$
\end{lemma}
{\bf Proof.} The proof of  Lemma \ref{l1} see \cite{Kallianpur}.

\begin{remark}\label{rem1} 
If the conditions of   Lemma \ref{l1} are valid, then there hold equalities
\begin{eqnarray}\label{jps1} 
E^P\{f_m|{\cal F}_k\}=f_k, \quad 0 \leq k \leq m, \quad m=\overline{1, \infty}, \quad P \in M.
\end{eqnarray}
\end{remark}

Let $f={\{f_{m}, {\cal F}_m \}_{m=0}^{\infty}}$ be a supermartingale relative to a convex set of equivalent measures $M$ and the filtration  ${\cal F}_m,\ m=\overline{0,\infty}.$ 
And let $G$ be a set of adapted non-decreasing  processes   $g={\{g_{m}\}_{m=0}^{\infty}}$, ${g_{0}=0},$ such that  $f+g={\{f_{m}+g_{m}\}_{m=0}^{\infty}}$ is a supermartingale concerning the family of measures    $M$ and the filtration   ${\cal F}_m,\ m=\overline{0,\infty}.$  

Introduce a partial ordering  $\preceq $ in the set of adapted non-decreasing processes $G.$

\begin{defin} We say that an adapted non-decreasing process ${g_{1}=\{g_{m}^{1}\}_{m=0}^{\infty}},$ ${g_{0}^{1}=0,} \ g_1 \in G,$  does not exeed an adapted non-decreasing process ${g_{2}=\{g_{m}^{2}\}_{m=0}^{\infty}},$ ${g_{0}^{2}=0}, \ g_{2} \in G,$  if   $P(g_{m}^{2}-g_{m}^{1}\geq 0)=1, \  m=\overline{1,\infty}.$ This partial ordering we denote by $g_{1}\preceq g_{2}.$
\end{defin}
For  every nonnegative adapted non-decreasing process $g=\{g_{m}\}_{m=0}^{\infty} \in G$ there exists limit  $\lim\limits_{m\to \infty}g_m$ which we denote by $g_{\infty}.$ 

\begin{lemma}\label{l2} Let  ${\tilde{G}}$ be a maximal chain in $G$ and for a certain  ${Q\in M}$ 
$\sup\limits_{g\in \tilde{G}}E_1^{Q}g=$ $\alpha^{Q}<\infty.$
Then there exists a sequence  $g^{s}=\{g_{m}^{s}\}_{m=0}^{\infty}\in \tilde G,$  ${s=1,2,...},$ such that
\begin{eqnarray*}
\sup\limits_{g\in \tilde{G}}E_1^{Q}g=\sup\limits_{s\geq 1}E_1^{Q}g^{s},
\end{eqnarray*}
where
$$ E_1^Qg=\sum\limits_{m=0}^\infty\frac{E^Q g_m}{2^m}, \quad g \in G.$$
\end{lemma}
{\bf Proof.}

Let  $0 < \varepsilon_s<\alpha^{Q}, \ s=\overline{1, \infty},$ be a sequence of real numbers satisfying conditions  $  \varepsilon_s > \varepsilon_{s+1}, \ \varepsilon_s \to 0,$ as $ s \to \infty.$ Then there exists an element  $g^s \in {\tilde{G}}$ such that
$\alpha^{Q} - \varepsilon_s < E_1^{Q}g^s \leq \alpha^{Q}, \  s=\overline{1,\infty}.$ The  sequence $g^s \in {\tilde{G}}, \  s=\overline{1,\infty},$ satisfies Lemma \ref{l2} conditions.

\begin{lemma}\label{l3} 
 If a  supermartingale  ${\{f_{m}, {\cal F}_m\}_{m=0}^{\infty}}$   relative to  a convex set of equivalent measures  $M$ is such that
\begin{eqnarray}\label{buti1}
 |f_{m}|\leq \varphi,  \quad m=\overline{0,\infty}, \quad E^Q\varphi < T < \infty, \quad   Q \in M, 
\end{eqnarray}
where a real number $T$ does not depend on  $Q \in M,$ then every maximal chain  ${\tilde{G}} \subseteq G$ contains  a maximal element.
\end{lemma}
{\bf Proof.}  Let  $g=\{g_m\}_{m=0}^\infty$ belong to  $G,$ then
\begin{eqnarray*}
 {E^{Q}(f_{m}+\varphi+g_{m})\leq f_{0}+T, \quad {m=\overline{1,\infty}}, \quad  Q\in M}.
\end{eqnarray*}
Then inequalities  ${f_{m}+\varphi \geq 0}, \ m=\overline{1,\infty},$ yield
 \begin{eqnarray*}
 {E^{Q}g_{m}\leq f_{0}+T}, \quad {m=\overline{1,\infty}},\quad {\{g_{m}\}_{m=0}^{\infty} \in G}.
 \end{eqnarray*}
Introduce for a certain $Q \in M$ an expectation  for  $g=\{g_m\}_{m=0}^\infty \in G$ 
\begin{eqnarray*}
E_1^Qg=\sum\limits_{m=0}^\infty\frac{E^Q g_m}{2^m}, \quad g \in G.
 \end{eqnarray*}
Let $\tilde G \subseteq G$ be a certain maximal chain.
Therefore, we have inequality
 \begin{eqnarray*}
 \sup\limits_{g\in \tilde{G}}E_1^{Q}g=\alpha_{0}^{Q}\leq f_{0}+T< \infty,
 \end{eqnarray*}
 where   $Q \in M$  and is fixed.
 Due to  Lemma \ref{l2},
\begin{eqnarray*}
{\sup\limits_{g\in \tilde{G}}E_1^{Q}g=\sup\limits_{s\geq 1}E_1^{Q}g^{s}}.
\end{eqnarray*}
In consequence of the linear ordering of elements of  ${\tilde{G}},$
\begin{eqnarray*}
{\max\limits_{1\leq s\leq k}g^{s}=g^{s_{0}(k)}}, \quad {1\leq s_{0}(k)\leq k},
\end{eqnarray*}
where  $s_{0}(k)$  is one of elements of the set  $\{1,2, \ldots, k\}$ on which the considered maximum is reached, that is, $1 \leq s_{0}(k) \leq k,$
and, moreover, 
\begin{eqnarray*}
 {g^{s_{0}(k)} \preceq g^{s_{0}(k+1)}}.
 \end{eqnarray*}
It is evident that
\begin{eqnarray*}
\max\limits_{1\leq s \leq k}E_1^Q g^s = E_1^Q g^{s_0(k)}. 
 \end{eqnarray*}
So, we obtain 
\begin{eqnarray*}
{\sup\limits_{s\geq 1}E_1^{Q}g^{s}=\lim\limits_{k\rightarrow\infty} \max\limits_{1\leq s\leq k}E_1^{Q}g^{s}=\lim\limits_{k\rightarrow\infty}E_1^{Q}g^{s_{0}(k)}=E_1^{Q}\lim\limits_{k\rightarrow\infty}g^{s_{0}(k)}=E_1^{Q}g^{0}},
\end{eqnarray*}
where ${g^{0}=\lim\limits_{k\rightarrow\infty}g^{s_{0}(k)}}$, and that there exists, due to monotony of   ${g^{s_{0}(k)}}$.
Thus,
\begin{eqnarray*}
{\sup\limits_{s\geq 1} E_1^{Q}g^{s}=E_1^{Q}g^{0}=\alpha_{0}^{Q}}.
\end{eqnarray*}
 Show that  $g^0={\{g_{m}^{0}\}_{m=0}^{\infty}}$ is  a maximal element in  ${\tilde G}$.
It is evident that $g^0$ belongs to $G.$
For every element   ${g=\{g_{m}\}_{m=0}^{\infty}} \in \tilde G$ two cases are possible:\\
1) ${\exists k}$ such that  ${g\preceq g^{s_{0}(k)}}$.\\
2) ${\forall k \quad g^{s_{0}(k)}\prec g}$.\\
In the first case  ${g\preceq g^{0}}.$
In the second one from 2) we have ${g^{0}\preceq g}$.
At the same time 

\begin{eqnarray}\label{05}
E_1^{Q}g^{s_{0}(k)}\leq E_1^{Q}g.
\end{eqnarray} 
By passing to the limit in   (\ref{05}), we obtain 
\begin{eqnarray}\label{06}
E_1^{Q}g^{0}\leq E_1^{Q}g.
\end{eqnarray} 
The strict inequality in   (\ref{06}) is impossible, since  ${E_1^{Q}g^{0}=\sup\limits_{g\in\tilde{G}}E_1^{Q}g}$.
Therefore, 
\begin{eqnarray}\label{07}
E_1^{Q}g^{0}= E_1^{Q}g.
\end{eqnarray} 
 The inequality  ${g^{0}\preceq g}$  and the equality (\ref{07}) imply that  ${g=g^{0}.}$

Let $M$ be a convex set of equivalent probability measures on   $\{\Omega, {\cal F}\}.$ Introduce into $M$ a metric  $|Q_1 -Q_2|=\sup\limits_{A \in {\cal F}}|Q_1(A) - Q_2(A)|,\  Q_1, \ Q_2 \in M.$ 

\begin{lemma}\label{qhon13}
Let $\{f_m, {\cal F}_m \}_{m=0}^\infty$  be a supermartingale relative to a compact convex set of equivalent measures  $M$ satisfying conditions (\ref{gon1}).
 If for every set of measures  $\{P_1, P_2, \ldots, P_s\}, \ s<\infty, \ P_i \in M, \ i=\overline{1,s},$
there exist a natural number   $1 \leq m_0<\infty,$  and depending on this
set of measures  ${\cal F}_{m_0-1}$ measurable nonnegative  random variable $\Delta_{m_0}^s,$   $P_1(\Delta_{m_0}^s>0)>0,$ satisfying conditions
\begin{eqnarray}\label{qhon14}
f_{m_0-1}-E^{ P_i}\{f_{m_0}|{\cal F}_{m_0-1}\} \geq \Delta_{m_0}^s, \quad i=\overline{1,s},  
\end{eqnarray}
then the set  $G$ of  adapted non-decreasing processes   $g=\{g_m\}_{m=0}^\infty , \  g_0=0,$ for which  $\{f_m+g_m\}_{m=0}^\infty$ is a supermartingale relative to the set of measures   $M$ contains nonzero element.
\end{lemma}
{\bf Proof.} 
For any point  $P_0 \in  M$ let us define a set of measures
\begin{eqnarray}\label{qd6}
M^{P_0, \bar \varepsilon_0}=\{Q \in  M, \ Q=(1-\alpha)P_0+\alpha P, \  P \in  M , \ 0 \leq \alpha \leq  \bar\varepsilon_0\},
\end{eqnarray} 
 $$\bar\varepsilon_0=\frac{L}{1+L}.$$
Prove that the set of measures  $M^{P_0, \bar \varepsilon_0}$ contains some ball of a positive  radius, that is, there exists a real number $\rho_0>0$ such that  
$M^{P_0, \bar \varepsilon_0}\supseteq C(P_0,\rho_0),$ where $C(P_0,\rho_0)=\{P\in  M,\ |P_0 - P |< \rho_0\}.$ 

Let  $C(P_0,\tilde \rho)=\{P\in  M, \ |P_0 - P |< \tilde \rho\}$ be an open ball in  $ M$ with the center at the point $P_0$ of a radius $0<\tilde \rho<1.$ Consider a map of the set $  M$ into itself given by the law: $f(P)=(1-\bar \varepsilon_0)P_0+\bar \varepsilon_0 P, \ P \in  M.$

The mapping  $ f(P)$   maps  an open ball  $C(P_2^{'},\delta)=\{P\in  M, |P_2^{'} - P |< \delta \}$  with the center at the point  $P_2^{'} $ of a radius  $\delta>0$  into an open ball with the center at the point  $(1- \bar \varepsilon_0)P_0+ \bar \varepsilon_0 P_2^{'}$ of the radius  $\bar \varepsilon_0\delta,$
since  $|(1-\bar \varepsilon_0)P_0+\bar \varepsilon_0 P_2^{'} - (1-\bar \varepsilon_0)P_0-\bar \varepsilon_0 P|=\bar \varepsilon_0|P_2^{'} - P|<\bar \varepsilon_0\delta.$ Therefore, an image of an open set  $ M_0 \subseteq  M$ is an open set $f(M_0)\subseteq  M,$ thus $ f(P)$ is an open mapping. Since $f(P_0)=P_0,$ then the image of the ball
 $C(P_0,\tilde \rho)=\{P\in  M,\ |P_0 - P |< \tilde\rho\}$ is a ball $C(P_0,\bar \varepsilon_0\tilde \rho)=\{P\in M,\ |P_0 - P |< \bar \varepsilon_0\tilde\rho\}$ and it is contained in  $f( M).$ 
 Thus, inclusions  $M^{P_0, \bar \varepsilon_0} \supseteq f( M) \supseteq C(P_0,\bar \varepsilon_0\tilde\rho)$ are valid.  Let us put $\bar \varepsilon_0\tilde\rho=\rho_0.$ Then we have $M^{P_0, \bar \varepsilon_0}\supseteq C(P_0,\rho_0),$ where  $C(P_0,\rho_0)=\{P\in  M,\ |P_0 - P |< \rho_0\}.$ 
Consider an open covering  $\bigcup\limits_{P_0 \in M}C(P_0, \rho_0)$ of the compact set  $ M.$  Due to compactness of  $ M,$ there exists a finite subcovering 
\begin{eqnarray}\label{qalmyd7}
 M=\bigcup\limits_{i=1}^v C(P_0^i, \rho_0) 
\end{eqnarray} 
with the center at  the points  $P_0^i \in M,  \ i=\overline{1, v}, $ and a covering by sets $ M^{P_0^i, \bar \varepsilon_0} \supseteq C(P_0^i, \rho_0),  \ i=\overline{1, v}, $
\begin{eqnarray}\label{qd7}
 M=\bigcup\limits_{i=1}^v M^{P_0^i, \bar \varepsilon_0}. 
\end{eqnarray} 
 
Consider the set of measures $P_0^i \in M,  \ i=\overline{1, v}. $
From  Lemma  \ref{qhon13} conditions, there exist a natural number   $1 \leq m_0<\infty,$  and depending on the
set of measures  $P_0^i \in M,  \ i=\overline{1, v}, $  ${\cal F}_{m_0-1}$ measurable nonnegative  random variable $\Delta_{m_0}^v,$  $P_0^1(\Delta_{m_0}^v>0)>0,$  such that
\begin{eqnarray}\label{cck1}
f_{m_0-1}-E^{ P_0^i}\{f_{m_0}|{\cal F}_{m_0-1}\} \geq \Delta_{m_0}^v, \quad i=\overline{1,v}.
\end{eqnarray}
 Due to  Theorem \ref{t1}, we have 
\begin{eqnarray}\label{ccj1}
f_{m_0-1} - E^Q\{f_{m_0}|{\cal F}_{m_0-1}\} \geq\frac{l}{1+L} \Delta_{m_0}^v=\varphi_{m_0}^v,   \quad Q \in M.
\end{eqnarray}
The last inequality imply 
\begin{eqnarray}\label{qk4}
E^Q\{f_{m_0-1}|{\cal F}_{s}\} - E^Q\{f_{m_0}|{\cal F}_{s}\} \geq E^{Q}\{ \varphi_{m_0}^v|{\cal F}_{s}\}, \quad Q \in M, \quad s <m_0.
\end{eqnarray}
But $ E^Q\{f_{m_0-1}|{\cal F}_{s}\} \leq f_s, \ s < m_0.$  Therefore,
\begin{eqnarray}\label{qk5}
 f_s - E^Q\{f_{m_0}|{\cal F}_{s}\} \geq E^{Q}\{\varphi_{m_0}^v|{\cal F}_{s}\}, \quad Q \in M, \quad s <m_0.
\end{eqnarray}
Since 
\begin{eqnarray}\label{qk6}
 f_{m_0} - E^Q\{f_{m}|{\cal F}_{m_0}\} \geq 0,  \quad Q \in M,  \quad m \geq m_0,
\end{eqnarray}
we have
\begin{eqnarray}\label{qk7}
E^Q\{ f_{m_0}|{\cal F}_{s}\} - E^Q\{f_{m}|{\cal F}_{s}\} \geq 0,  \quad Q \in M, \quad s <m_0, \quad m \geq m_0.
\end{eqnarray}
Adding  (\ref{qk7}) to (\ref{qk5}), we obtain
\begin{eqnarray}\label{qk8}
  f_s  - E^Q\{f_{m}|{\cal F}_{s}\}\geq E^{Q}\{ \varphi_{m_0}^v|{\cal F}_{s}\},  \quad Q \in M, \quad s <m_0, \quad m \geq m_0,
\end{eqnarray}
or
\begin{eqnarray}\label{qk9}
  f_s  - E^Q\{f_{m}|{\cal F}_{s}\}\geq E^{Q}\{\varphi_{m_0}^v|{\cal F}_{s}\}\chi_{[m_0,\infty)}(m) -  \varphi_{m_0}^v\chi_{[m_0,\infty)}(s), 
\end{eqnarray}
$$   Q \in M, \quad s \leq m_0, \quad m \geq m_0.$$
Introduce an adapted non-decreasing process
\begin{eqnarray*}
g^{m_0}=\{g_m^{m_0}\}_{m=0}^\infty, \quad g_m^{m_0}= \varphi_{m_0}^v\chi_{[m_0,\infty)}(m),
\end{eqnarray*}
where $\chi_{[m_0,\infty)}(m)$ is an indicator function of the set $ [m_0,\infty). $
Then   (\ref{qk9}) implies  that
\begin{eqnarray*}
E^Q\{ f_m+ g_m^{m_0}|{\cal F}_k\}\leq f_k +g_k^{m_0}, \quad  0 \leq k \leq m, \quad Q \in M.
\end{eqnarray*}

 In the Theorem \ref{ctt5} a convex set of equivalent  measures 
\begin{eqnarray}\label{self100}
M =\{Q, \ Q=\sum\limits_{i=1}^n \alpha_iP_i, \  \alpha_i \geq 0, \ i=\overline{1, n}, \ \sum\limits_{i=1}^n \alpha_i=1\}
\end{eqnarray}
satisfies conditions
\begin{eqnarray}\label{self101}
0<l \leq \frac{dP_i}{dP_j} \leq L < \infty,\quad  i,j=\overline{1,n},
\end{eqnarray} 
where  $l,\ L $  are real  numbers.

 Denote by  $G$  the set of all  adapted non-decreasing processes  $g=\{g_m\}_{m=0}^{\infty},$ $ g_0=0,$ such that  $\{f_m+g_m\}_{m=0}^{\infty}$ is a supermartingale relative to all measures from $M.$
\begin{thm}\label{ctt5} 
Let   a supermartingale ${\{f_{m},  {\cal F}_m\}_{m=0}^{\infty}}$ relative to the set of measures (\ref{self100})   satisfy the conditions (\ref{buti1}),  and let  there exist a natural number $1 \leq m_0<\infty,$ and ${\cal F}_{m_0-1}$ measurable nonnegative  random value $\varphi_{m_0}^n,$  $P_1(\varphi_{m_0}^n>0)>0,$ such that
\begin{eqnarray}\label{cqkgon14}
f_{m_0-1}-E^{ P_i}\{f_{m_0}|{\cal F}_{m_0-1}\} \geq \varphi_{m_0}^n, \quad i=\overline{1,n}.
\end{eqnarray}
If  for  the maximal  element 
  $g^{0}=\{g_m^{0}\}_{m=0}^{\infty}$ in a certain  maximal chain  $\tilde G \subseteq G$ the  equalities  
\begin{eqnarray}\label{c5}
E^{P_i}(f_{\infty}+g_{\infty}^{0})=f_{0}, \quad P_i \in M,  \quad i=\overline{1,n},
\end{eqnarray} 
are valid, where $f_{\infty}=\lim\limits_{m \to \infty}f_m,$ $g_{\infty}^0=\lim\limits_{m \to \infty}g_m^0,$ then there hold equalities 
\begin{eqnarray}\label{c5c}
E^{P}\{f_{m}+g_{m}^{0}|{\cal F}_k\}=f_{k}+g_{k}^{0}, \quad 0 \leq k \leq m, \quad  m=\overline{1, \infty}, \quad P \in M. 
\end{eqnarray} 
\end{thm}
{\bf Proof.} 
The set $M$ is compact one  in the introduced metric topology. From the inequalities  (\ref{cqkgon14})  and the formula
\begin{eqnarray}\label{ns23}
 E^{Q}\{f_{m_0}|{\cal F}_{m_0-1}\}=\frac{\sum\limits_{i=1}^{n}\alpha_i E^{P_1}\{\varphi_i|{\cal F}_{m_0-1}\}E^{P_i}\{f_{m_0}|{\cal F}_{m_0-1}\}}{\sum\limits_{i=1}^{n}\alpha_i E^{P_1}\{\varphi_i|{\cal F}_{m_0-1}\}}, \quad Q \in M,
\end{eqnarray}
where $\varphi_i=\frac{dP_i}{dP_1},$ we obtain
\begin{eqnarray}\label{dargon14}
f_{m_0-1}-E^{Q}\{f_{m_0}|{\cal F}_{m_0-1}\} \geq \varphi_{m_0}^n, \quad Q \in M.
\end{eqnarray}
The inequalities (\ref{self101}) lead to inequalities
\begin{eqnarray}\label{mazgon14}
\frac{1}{nL} \leq \frac{dQ}{dP} \leq n L, \quad P, Q \in M.
\end{eqnarray}
  Inequalities (\ref{dargon14}) and (\ref{mazgon14}) imply that  conditions of  Lemma \ref{qhon13}  are satisfied  for any set of measures $Q_1,\ldots,Q_s \in M.$ Hence, it follows  that the set  $G$   contains nonzero element. Let $\tilde G \subseteq G$ be a maximal chain in  $ G$ satisfying condition of  Theorem \ref{ctt5}.
Denote by  $g^0=\{g_m^0\}_{m=0}^\infty,\ g_0^0=0,$ a maximal element  in $\tilde G \subseteq G.$ 
  Theorem \ref{ctt5} and Lemma  \ref{l3} yield  that as $\{f_m\}_{m=0}^{\infty}$ and $\{g_m^0\}_{m=0}^\infty$ are uniformly integrable relative to each measure from $M.$ There exist   therefore limits
$$ \lim\limits_{m \to \infty}f_m=f_{\infty}, \quad \lim\limits_{m \to \infty}g_m^0=g_{\infty}^0$$
with probability 1.
Due to  Theorem  \ref{ctt5}  condition,  in this  maximal chain 
$$E^{P_i}(f_{\infty}+g_{\infty}^0)=f_0, \quad P_i \in M, \quad  i=\overline{1,n} .  $$
Since $ \{f_m+g_m^0\}_{m=0}^\infty$ is a supermartingale concerning all measures from $M,$ we have
\begin{eqnarray}\label{cmykald8}
 E^{P_i}(f_{m}+g_m^0) \leq   E^{P_i}(f_{k}+g_k^0)\leq f_0, \quad k< m, \quad m=\overline{1,\infty}, \quad  P_i \in M.
\end{eqnarray}
By passing to the limit in (\ref{cmykald8}), as $m \to \infty,$ we obtain
\begin{eqnarray}\label{cmykald7} 
f_{0}=E^{P_i}(f_{\infty}+g_{\infty}^0) \leq  E^{P_i}(f_{k}+g_k^0)\leq f_0,  \quad k= \overline{1,\infty}, \quad  P_i \in M. 
\end{eqnarray}
So, $ E^{P_i}(f_{k}+g_k^0)= f_0, \ k=\overline{1,\infty}, \  P_i \in M, \ i=\overline{1,n}.$
Taking into account Remark \ref{rem1} we have
\begin{eqnarray}\label{cd8}
 E^{P_i}\{f_{m}+g_m^0|{\cal F}_k\}=f_k+ g_k^0, \quad 0 \leq  k \leq m, \ m=\overline{1,\infty},
\end{eqnarray}
 $$ P_i \in M, \ i=\overline{1,n}.$$
Hence,
$$E^{P}\{f_{m}+g_m^0|{\cal F}_{k}\}=$$
\begin{eqnarray}\label{ck1}
 \frac{\sum\limits_{i=1}^{n}\alpha_i E^{P_1}\{\varphi_i|{\cal F}_{k}\}E^{P_i}\{f_{m}+g_m^0|{\cal F}_{k}\}}{\sum\limits_{i=1}^{n}\alpha_i E^{P_1}\{\varphi_i|{\cal F}_{k}\}}=f_k+g_k^0,  \quad 0 \leq k \leq m, 
\end{eqnarray}
$$ P \in M,$$
where $ \varphi_i=\frac{dP_i}{dP_1}, \ i=\overline{1,n}.$ Theorem  \ref{ctt5} is proved.
 
Let $M$ be a convex set of equivalent measures. Below,  $G_s$ is  a set  of adapted non-decreasing  processes $\{g_m\}_{m=0}^\infty,$ $ \ g_0=0,$ for which $\{f_m+g_m\}_{m=0}^\infty$ is a supermartingale relative
to all measures from 
 \begin{eqnarray}\label{Wov3}
\hat M_s=\{Q, Q=\sum\limits_{i=1}^s\gamma_i \hat P_i, \ \gamma_i \geq 0, \ i=\overline{1,s}, \ \sum\limits_{i=1}^s\gamma_i=1\},
\end{eqnarray}
where  $\hat P_1, \ldots,\hat P_s \in M$ and satisfy conditions
 \begin{eqnarray}\label{fifa1}
0< {l\leq\frac{d\hat P_i}{d\hat P_j}\leq L}< \infty,\quad  i,j=\overline{1,s},  
\end{eqnarray}
 $l, L$ are  real numbers depending on the set of measures  $\hat P_1, \ldots,\hat P_s \in M.$

\begin{defin}\label{1000h} 
Let   a  supermartingale   $\{f_m, {\cal F}_m\}_{m=0}^\infty$ relative to  a convex set of equivalent measures  $M $ satisfy  conditions (\ref{buti1}). 
 We  call it  regular one if for every   set of measures (\ref{Wov3}) satisfying conditions (\ref{fifa1}) there exist a natural number $1 \leq m_0<\infty,$ and   ${\cal F}_{m_0-1}$ measurable nonnegative random value $\varphi_{m_0}^s,$ $\hat P_1(\varphi_{m_0}^s>0)>0, $ such that  the inequalities 
$$f_{m_0-1}- E^{\hat P_{i}}\{f_{m_0}|{\cal F}_{m_0-1}\}\geq \varphi_{m_0}^s, \quad i=\overline{1,s},$$
 hold and for the maximal element $g^s=\{g_m^s\}_{m=0}^\infty$
in a certain  maximal chain $\tilde G_s \subseteq G_s$  the equalities
\begin{eqnarray}\label{Wov1}
E^{\hat P_i}\{f_m +g_m^s|{\cal F}_{k}\}=f_{k}+g_{k}^s, \quad 0 \leq  k \leq m, \quad i=\overline{1,s},\quad m=\overline{1, \infty},
\end{eqnarray} 
are valid. Moreover, there exists  an adapted nonnegative process $\bar g^0=\{\bar g_m^0\}_{m=0}^\infty, $ $ \bar g_0^0=0,$  $ E^P \bar g_m^0<\infty, \  m=\overline{1, \infty}, \  P \in M,$ not depending on the set of measures 
$\hat P_1, \ldots,\hat P_s$  such that
\begin{eqnarray}\label{WOv2}
E^{\hat P_i}\{g_m^s- g_{m-1}^s|{\cal F}_{m-1}\}=E^{\hat P_i}\{\bar g_m^0|{\cal F}_{m-1}\}, \quad m=\overline{1,\infty}, \quad i=\overline{1,s}.
\end{eqnarray}
\end{defin}
The next Theorem describes  regular supermartingales.
\begin{thm}\label{ct4}   
  Let $\{f_m, {\cal F}_m\}_{m=0}^\infty$ be a regular   supermartingale  relative to  a convex set of equivalent measures  $M. $  
Then for the maximal element   $g^0=\{g_m^0\}_{m=0}^{\infty}$ in  a certain  maximal chain  $\tilde G \subseteq  G$ the equalities  
\begin{eqnarray*}
E^{P_0}(f_m +g_m^0)=f_0, \quad m=\overline{1,\infty},  \quad P_0 \in M,
\end{eqnarray*} 
are valid. There exists   a martingale   $\{\bar M_m,{\cal F}_m\}_{m=0}^\infty$  relative to the family of measures   $M$  such that 
\begin{eqnarray*}
 f_m=\bar M_m- g_m^0, \quad m=\overline{1,\infty}. 
\end{eqnarray*}
Moreover, for the martingale  $\{\bar M_m,{\cal F}_m\}_{m=0}^\infty$   the representation 
\begin{eqnarray*}
\bar M_m=E^{P_0}\{f_{\infty}+ g_{\infty}|{\cal F}_m\},\quad m=\overline{1,\infty},  \quad P_0 \in M,
\end{eqnarray*}
holds, where  $ f_{\infty}+ g_{\infty}=\lim\limits_{m \to \infty}(f_m+g_m).$
\end{thm} 
{\bf Proof.}   For any finite set of measures $P_1, \ldots,P_n,$ $P_i \in M, \ i=\overline{1,n},$
let us introduce into consideration two sets of measures 
\begin{eqnarray*}
 M_n=\{P, \ P=\sum\limits_{i=1}^n\alpha_iP_i, \  \alpha_i \geq 0, \ i=\overline{1,n}, \ \sum\limits_{i=1}^n\alpha_i=1\},
\end{eqnarray*}
\begin{eqnarray*}
\tilde M_n=\{P, \  P=\sum\limits_{i=1}^n\alpha_iP_i, \  \alpha_i > 0, \  i=\overline{1,n}, \ \sum\limits_{i=1}^n\alpha_i=1\}.
\end{eqnarray*}
Let $\hat P_1, \ldots, \hat P_s$ be a certain subset of measures    from $\tilde M_n.$  For every measure $\hat P_i \in \tilde M_n$ the representation   
 $\hat P_i=\sum\limits_{k=1}^n\alpha_k^iP_k $ is valid, where $\alpha_k^i>0,  \ i=\overline{1,s}, \ k=\overline{1,n}.$   The representation for $\hat P_i, \ i=\overline{1,s}, $ imply the validity  of inequalities
\begin{eqnarray*}
0<l=\min\limits_{i,j}\frac{\min\limits_{k}\alpha_k^i}{\max\limits_{k}\alpha_k^j} \leq \frac{d\hat P_i}{d\hat P_j}\leq \max\limits_{i,j}\frac{\max\limits_{k}\alpha_k^i}{\min\limits_{k}\alpha_k^j}=L < \infty, \quad i,j=\overline{1,s}.
\end{eqnarray*}
 Denote by $G_s$  a set  of adapted non-decreasing  processes $\{g_m\}_{m=0}^\infty, \ g_0=0,$ for which $\{f_m+g_m\}_{m=0}^\infty$ is a supermartingale relative
to all measures from 
  $$\hat M_s=\{Q, \  Q=\sum\limits_{i=1}^s\gamma_i \hat P_i, \ \gamma_i \geq 0, \ i=\overline{1,s}, \ \sum\limits_{i=1}^s\gamma_i=1\}.$$
In accordance with the definion of a regular supermartingale,  there exist a natural number $1 \leq  m_0<\infty,$ and   ${\cal F}_{m_0-1}$ measurable nonnegative random value $\varphi_{m_0}^s,$ $\hat P_1(\varphi_{m_0}^s>0)>0, $  such that the inequalities there hold  
$$f_{m_0-1}- E^{\hat P_{i}}\{f_{m_0}|{\cal F}_{m_0-1}\}\geq \varphi_{m_0}^s, \quad i=\overline{1,s},$$
and for a maximal element $g^s=\{g_m^s\}_{m=0}^\infty$
in a certain maximal chain $\tilde G_s \subseteq G_s$ there hold equalities (\ref{Wov1}), (\ref{WOv2}). Equalities (\ref{WOv2}) yield the equalities
$$E^{Q}\{g_{m}^s -  g_{m-1}^s|{\cal F}_{m-1}\}=$$
\begin{eqnarray*}\label{0ck1}
 \frac{\sum\limits_{i=1}^{s}\gamma_i E^{\hat P_1}\{\hat \varphi_i|{\cal F}_{m-1}\}E^{\hat P_i}\{g_{m}^s - g_{m-1}^s|{\cal F}_{m-1}\}}{\sum\limits_{i=1}^{s}\gamma_i E^{\hat P_1}\{\hat \varphi_i|{\cal F}_{m-1}\}}=   
\end{eqnarray*}
\begin{eqnarray}\label{Wau1}
\frac{\sum\limits_{i=1}^{s}\gamma_i E^{\hat P_1}\{\hat \varphi_i|{\cal F}_{m-1}\}E^{\hat P_i}\{\bar g_{m}^0|{\cal F}_{m-1}\}}{\sum\limits_{i=1}^{s}\gamma_i E^{\hat P_1}\{\hat \varphi_i|{\cal F}_{m-1}\}}=E^{Q}\{\bar g_{m}^0|{\cal F}_{m-1}\}, 
\end{eqnarray}
$$  m=\overline{1, \infty}, \quad Q \in \hat M_s.$$
where $\hat \varphi_i=\frac{d\hat P_i}{d\hat P_1}, \ i=\overline{1,n}.$
Taking into account the equalities (\ref{Wov1}), we obtain
$$E^{Q}\{f_m+g_{m}^s |{\cal F}_{m-1}\}=$$
\begin{eqnarray*}\label{Fck1}
 \frac{\sum\limits_{i=1}^{s}\gamma_i E^{\hat P_1}\{\hat \varphi_i|{\cal F}_{m-1}\}E^{\hat P_i}\{f_m+g_{m}^s|{\cal F}_{m-1}\}}{\sum\limits_{i=1}^{s}\gamma_i E^{\hat P_1}\{\hat \varphi_i|{\cal F}_{m-1}\}}=   
\end{eqnarray*}
\begin{eqnarray}\label{FWau1}
f_{m-1}+g_{m-1}^s, \quad  m=\overline{1, \infty}, \quad Q \in \hat M_s.
\end{eqnarray}

Thus, we have
\begin{eqnarray}\label{Wau2}
E^{Q}\{g_{m}^s -  g_{m-1}^s|{\cal F}_{m-1}\}=E^{Q}\{\bar g_{m}^0|{\cal F}_{m-1}\},   \quad m=\overline{1, \infty},   \quad Q \in \hat M_s.
\end{eqnarray}
\begin{eqnarray}\label{FWau2}
E^{Q}\{f_m+g_{m}^s|{\cal F}_{m-1}\}=f_{m-1}+g_{m-1}^s,   \quad m=\overline{1, \infty},   \quad Q \in \hat M_s.
\end{eqnarray}
Let us introduce into consideration a random process $\{N_m, {\cal F}_m\}_{m=0}^\infty,$
 where
$$ N_0=f_0, \quad N_m=f_m+\sum\limits_{i=1}^m \bar g_m^0, \quad m=\overline{1, \infty}.$$
It is evident that $E^{Q}|N_m|< \infty, \ m=\overline{1, \infty}, \ Q \in  \hat M_s.$ The definition of  $\{N_m, {\cal F}_m\}_{m=0}^\infty$ and the formulae (\ref{Wau2}), ( \ref{FWau2}) yield
$$E^{Q}\{N_{m-1}- N_m |{\cal F}_{m-1}\}=
E^{Q}\{f_{m-1} - f_m - \bar g_m^0|{\cal F}_{m-1}\}=$$ $$=E^{Q}\{g_m^s- g_{m-1}^s -\bar g_m^0|{\cal F}_{m-1}\}=0, \quad m=\overline{1, \infty}, \quad \ Q \in  \hat M_s.$$
The last equalities imply 
$$ E^{Q}\{N_m|{\cal F}_{m-1}\}=N_{m-1}, \quad m=\overline{1, \infty},  \quad Q \in \hat M_s.$$
Due to arbitrariness of the set of measures $\hat P_1, \ldots, \hat P_s, $ $ \hat P_i \in \tilde M_n,$ we have
\begin{eqnarray}\label{wy1}
E^{P}\{N_m|{\cal F}_{m-1}\}=N_{m-1}, \quad  P \in \tilde M_n,\quad m=\overline{1, \infty}.
\end{eqnarray}
So,  the set $G_0$   of  adapted non-decreasing processes $\{g_m\}_{m=0}^\infty,$ $ \ g_0=0,$ for which $\{f_m+g_m\}_{m=0}^\infty$ is a supermartingale relative
to all measures from $\tilde M_n$ contains nonzero element $\tilde g^0=\{\tilde g_m^0\}_{m=0}^\infty, \ \tilde g_0^0=0,$ $\tilde g_m^0=\sum\limits_{i=1}^m \bar g_m^0, \ m=\overline{1, \infty},$ which is  a maximal element in a  maximal chain $\tilde G_0$ containing this element. Really, if  $g^0=\{ g_m^0\}_{m=0}^\infty, \  g_0^0=0,$ is a maximal element 
 in the maximal chain $\tilde G_0 \subseteq G_0,$ then there hold inequalities
\begin{eqnarray}\label{my1}
E^{P_0}\{f_m+ g_m^0|{\cal F}_{k}\} \leq f_{k}+ g_{k}^0,\quad m=\overline{1,\infty},  \quad 1 \leq k \leq m,\quad P_0 \in \tilde M_n,
\end{eqnarray}
\begin{eqnarray}\label{my2}
E^{P_0}(f_m+ g_m^{0})\leq f_0 ,\quad m=\overline{1,\infty},  \quad P_0 \in \tilde M_n.
\end{eqnarray}
and inequality $\tilde g^0 \preceq g^0$  meaning that $\tilde g_m^{0} \leq  g_m^{0}, \ m=\overline{0, \infty}.$
Equalities (\ref{wy1}) yield
\begin{eqnarray}\label{ty2}
E^{P_0}(f_m+\tilde g_m^{0})= f_0 ,\quad m=\overline{1,\infty},  \quad P_0 \in \tilde M_n.
\end{eqnarray}
 Inequalities  (\ref{my2})  and equalities  (\ref{ty2}) imply
\begin{eqnarray}\label{ty3}
f_0 \geq E^{P_0}(f_m+ g_m^{0})\geq  E^{P_0}(f_m+\tilde g_m^{0})= f_0 ,\quad m=\overline{1,\infty},  \quad P_0 \in \tilde M_n.
\end{eqnarray}
The last inequalities lead to equalities
\begin{eqnarray}\label{ty4}
 E^{P_0}( g_m^{0}- \tilde g_m^{0})= 0, \quad m=\overline{1,\infty},  \quad P_0 \in \tilde M_n.
\end{eqnarray}
But
\begin{eqnarray}\label{ty5}
 g_m^{0}- \tilde g_m^{0}\geq 0,\quad m=\overline{0,\infty}.
\end{eqnarray}
The equalities (\ref{ty4}) and inequalities (\ref{ty5}) yield  $g_m^{0}=\tilde g_m^{0}, \ m=\overline{0,\infty},$ or $\tilde g^0=g^0.$

Prove that $G_n=G_0,$ where $G_n$ is a set of non-decreasing processes  $g=\{g_m\}_{m=0}^{\infty}$ such that  $\{f_m+g_m\}_{m=0}^{\infty}$  is a supermartingale relative to all measures from  $M_n.$  Really, if  $g=\{g_m\}_{m=0}^{\infty}$  is a non-decreasing process from  $G_n,$ then it belongs to $G_0,$ owing to that   $M_n \supset \tilde M_n$ and   $G_n \subseteq G_0.$
Suppose that  $g=\{g_m\}_{m=0}^{\infty}, \ g_0=0,$  is a non-decreasing process from  $G_0.$ It means that 
\begin{eqnarray}\label{c28}
 E^Q\{f_m+  g_m|{\cal F}_{k}\}\leq f_{k}+g_{k}, \quad m=\overline{1,\infty},  \quad 0 \leq  k\leq m, \quad Q \in \tilde M_n.
\end{eqnarray} 
The last inequalities can be written in the form 
$$\sum\limits_{i=1}^n\alpha_i\int\limits_{A}(f_m+  g_m)dP_i\leq \sum\limits_{i=1}^n\alpha_i\int\limits_{A}(f_{k}+  g_{k})dP_i, \quad  m=\overline{1,\infty},  \quad  0 \leq  k \leq m, $$ 
$$   A \in {\cal F}_{k}, \quad  \alpha_i>0, \quad i=\overline{1,n}. $$
By passing to the limit, as $\alpha_j \to 0,  \ \alpha_j>0, \ j \neq i, \  \alpha_i \to 1,$ we obtain 
$$\int\limits_{A}(f_m+  g_m)dP_i\leq \int\limits_{A}(f_{k}+  g_{k})dP_i, \quad i=\overline{1,n},  \quad A \in {\cal F}_{k}, \quad  m=\overline{1,\infty},  \quad  0 \leq  k\leq m.$$
The last inequalities yield inequalities
$$\sum\limits_{i=1}^n\alpha_i\int\limits_{A}(f_m+  g_m)dP_i\leq \sum\limits_{i=1}^n\alpha_i\int\limits_{A}(f_{k}+  g_{k})dP_i, \quad  m=\overline{1,\infty}, \quad   0 \leq k\leq m,  $$ $$  A \in {\cal F}_{k},  \quad  \alpha_i \geq 0, \quad i=\overline{1,n},$$
or 
\begin{eqnarray*}
 E^Q\{f_m+  g_m|{\cal F}_{k}\}\leq f_{k}+g_{k}, \quad m=\overline{1,\infty},  \quad  0 \leq k\leq m,  \quad Q \in M_n.
\end{eqnarray*} 
It means that  $g=\{g_m\}_{m=0}^{\infty}$ belongs to  $G_n.$  On the basis of the above proved,  for the maximal element $ \tilde g^0=\{ \tilde g_m^0\}_{m=0}^{\infty}$  in   the maximal chain   $\tilde G_0 \subseteq G_0$ the equalities 
\begin{eqnarray}\label{clobal} 
 E^Q \{f_m+ \tilde g_m^0|{\cal F}_{k}\}=f_{k}+ \tilde g_{k}^0, \quad  m=\overline{1,\infty},  \quad 1 \leq k\leq m,  \quad Q \in \tilde M_n,
\end{eqnarray}
\begin{eqnarray}\label{cchukal} 
 E^Q (f_m+ \tilde g_m^0)=f_0, \quad  m=\overline{1,\infty},   \quad Q \in \tilde M_n,
\end{eqnarray}
are valid.  From proved equality   $G_n=G_0,$  it follows that  $\tilde G_0$  is a maximal chain in  $G_n .$ 

As far as, $G_0$ coincides with $G_n$  we proved that  the maximal element $\tilde g^0$ in a certain maximal chain in  $G_n$  satisfies equalities
\begin{eqnarray}\label{my5}
E^{P_0}\{f_m+\tilde g_m^0|{\cal F}_{k}\} = f_{k}+\tilde g_{k}^0,\quad m=\overline{1,\infty},  \quad 1 \leq k \leq m,\quad P_0 \in  M_n,
\end{eqnarray}
\begin{eqnarray}\label{my6}
E^{P_0}(f_m+\tilde g_m^{0})=f_0 ,\quad  m=\overline{1,\infty},  \quad P_0 \in M_n.
\end{eqnarray}
Due to arbitrariness of the set of measure $P_1, \ldots, P_n,\ P_i \in M,$  the set $G$ contains nonzero element $\tilde g^0$ and in the maximal chain   $\tilde G \subseteq G$ containing element $\tilde g^0$ the  maximal element $g^0=\{g_m^0\}_{m=0}^{\infty}, g_0^0=0,$ coincides with $\tilde g^0.$ The last statement can be proved as in the case of maximal chain $\tilde G_0.$
So,
\begin{eqnarray}\label{my3}
E^{P_0}\{f_m+g_m^0|{\cal F}_{k}\} = f_{k}+g_{k}^0,\quad m=\overline{1,\infty},  \quad 1 \leq k \leq m,\quad P_0 \in M,
\end{eqnarray}
\begin{eqnarray}\label{my4}
E^{P_0}(f_m+g_m^{0})=f_0 ,\quad m=\overline{1,\infty},  \quad P_0 \in M.
\end{eqnarray}
Denote by  $\{\bar M_m, {\cal F}_m\}_{m=0}^{\infty}$ a martingale relative to the set of measures $M,$ where   $\bar M_m=f_m+g_m^{0}, \ m=\overline{1,\infty}.$
Due to  Theorem \ref{ct4} conditions, the supermartingale    $\{f_m, {\cal F}_m\}_{m=0}^\infty$
and non-decreasing process      $g^0=\{g_m^0\}_{m=0}^\infty$ are uniformly integrable relative to any measure from  $M,$  since for the non-decreasing process     $g^0=\{g_m^0\}_{m=0}^\infty$ there hold bounds    $E^Pg_{m}^0< T+f_0, \ m=\overline{1,\infty}, \ P \in M.$
Therefore, the martingale   $\{\bar M_m, {\cal F}_m\}_{m=0}^\infty$ is uniformly integrable  relative to any measure from   $M.$  
So, with probability 1 relative to every measure from   $M$ there exist limits  
\begin{eqnarray*}
\lim\limits_{m \to \infty} \bar M_{m}=M_{\infty}= f_{\infty}+ g_{\infty}^0, \quad \lim\limits_{m \to \infty}f_{m} = f_{\infty}, \quad \lim\limits_{m \to \infty}g_{m}^0 = g_{\infty}^0.
\end{eqnarray*}
Moreover, the representation   
\begin{eqnarray}\label{cp39}
\bar  M_m=E^P\{( f_{\infty}+ g_{\infty}^0)| {\cal F}_m\}, \quad m=\overline{1,\infty},   \quad P \in M,
\end{eqnarray}
 holds, where
  $\bar M=\{ \bar M_m\}_{m=0}^{\infty}$  does not depend on  $P \in M.$
 
In the next theorem we give the necessary and sufficient conditions of regularity of supermartingales.
\begin{thm}\label{reww1}
Let  a supermartingale    $\{f_{m},\ {\cal F}_m\}_{m=0}^{\infty}$ relative to a convex set of equivalent measures $M$ satisfy conditions (\ref{buti1}).
 The necessary and sufficient conditions for it to  be a regular one is the existence 
of adapted nonnegative random process $\bar g^0=\{\bar g_m^0\}_{m=0}^\infty, \ \bar g_0^0=0,$ $E^P\bar g_m^0<\infty,\ m=\overline{1, \infty}, \ P \in M, $ such that  equalities
\begin{eqnarray}\label{reww2}
E^P\{f_{m-1} -f_m|{\cal F}_{m-1}\}=E^P\{\bar g_m^0|{\cal F}_{m-1}\}, \quad m=\overline{1, \infty} , \quad P \in M,
\end{eqnarray} 
are valid.
\end{thm}
{\bf Proof.} 
{ \bf Necessity.} If   $\{f_{m},\ {\cal F}_m\}_{m=0}^{\infty}$ is  a regular supermartingale,  then there exist a martingale  $\{\bar M_{m},\ {\cal F}_m\}_{m=0}^{\infty}$ and a non-decreasing nonnegative random process  $\{g_{m},\ {\cal F}_m\}_{m=0}^{\infty},$ $ \ g_0=0,$ such that
\begin{eqnarray}\label{reww3}
f_m = \bar M_m - g_m, \quad m=\overline{1, \infty}.
\end{eqnarray}
As before, equalities (\ref{reww3}) yield inequalities $E^Pg_m \leq f_0+T, \  m=\overline{1, \infty},$
and  equalities
$$E^P\{f_{m-1} -f_m|{\cal F}_{m-1}\}=$$
\begin{eqnarray}\label{reww4}
=E^P\{ g_m - g_{m-1}|{\cal F}_{m-1}\}=E^P\{\bar g_m^0|{\cal F}_{m-1}\}, \quad m=\overline{1, \infty} , \quad P \in M,
\end{eqnarray}
where we introduced the denotation $\bar g_m^0=g_m - g_{m-1} \geq 0.$
It is evident that $E^P\bar g_m^0\leq 2( f_0+T).$

{\bf Sufficiency.}  If there exists an adapted nonnegative random process $\bar g^0=\{\bar g_m^0\}_{m=0}^\infty, \ \bar g_0^0=0,$  $ E^P\bar g_m^0<\infty, \ m=\overline{1, \infty}, $ such that the equalities (\ref{reww2}) are valid, then let us consider a  random process $\{\bar M_{m},\ {\cal F}_m\}_{m=0}^{\infty},$ where
\begin{eqnarray}\label{reww5}
\bar M_0=f_0, \quad \bar M_m=f_m+\sum\limits_{i=1}^m\bar g_m^0, \quad m=\overline{1, \infty}.
\end{eqnarray}
It is evident that $E^P|\bar M_m|< \infty$ and
$$E^P\{\bar M_{m-1} -  \bar M_m|{\cal F}_{m-1}\}=E^P\{f_{m-1} - f_m- \bar g_m^0|{\cal F}_{m-1}\}=0.$$
Theorem \ref{reww1} is proven.

In the next Theorem we describe  the structure of non-decreasing process for a regular supermartingale.
\begin{thm}\label{kj1}   
 Let  a  supermartingale $\{f_m, {\cal F}_m\}_{m=0}^\infty$  relative to a convex set of equivalent measures $M$  satisfy conditions (\ref{buti1}). The necessary and sufficient conditions for it to be  regular one is the existence  of a non-decreasing  adapted  process $ g=\{ g_m\}_{m=0}^{\infty}, \ g_0=0, $ and   adapted  processes
$\bar \Psi^{j}=\{ \bar \Psi^{j}_m\}_{m=0}^{\infty}, $ $  \bar \Psi^{j}_0=0, \ j=\overline{1,n},$ such that between elements $g_m, \ m=\overline{1,\infty},$  of non-decreasing process  $g=\{g_m\}_{m=0}^{\infty}$     the relations
\begin{eqnarray}\label{kj2}
g_m -g_{m-1}=f_{m-1}- E^{P_j}\{f_m|{\cal F}_{m-1}\}+  \bar \Psi^{j}_m, \quad  m=\overline{1,\infty}, \quad  j=\overline{1,n},
\end{eqnarray}
 are valid for each  set of measures $P_1, \ldots,P_n \in M$, where  $E^{P_j}|\bar \Psi^{j}_m|< \infty, $  $E^{P_j}\{\bar \Psi^{j}_m|{\cal F}_{m-1}\}=0, \ j=\overline{1,n}, \ m=\overline{1,\infty}. $  
\end{thm}
{\bf Proof.} {\bf The necessity.} 
Let $\{f_m, {\cal F}_m\}_{m=0}^\infty$ be a regular supermartingale. Then for it
 the representation 
\begin{eqnarray}\label{kkj3}
 f_m+g_m=M_m, \quad  m=\overline{1,\infty}, \quad  j=\overline{1,n},
\end{eqnarray}
is valid,  where $ \{g_m\}_{m=0}^\infty, \ g_0=0,$ is a non-decreasing adapted process, \\ $\{M_m, {\cal F}_m\}_{m=0}^\infty$ is a martingale relative to the set of measures $M.$
 For any finite set of measures $P_1, \ldots,P_n \in M,$ we have
\begin{eqnarray}\label{kj3}
 E^{P_j}\{f_m+g_m|{\cal F}_{m-1}\}=f_{m-1}+g_{m-1}, \quad  m=\overline{1,\infty}, \quad  j=\overline{1,n}.
\end{eqnarray}
Hence, we have
\begin{eqnarray}\label{kj4}
 E^{P_j}\{g_m- g_{m-1}|{\cal F}_{m-1}\}=
\end{eqnarray}
$$ f_{m-1}- E^{P_j}\{f_m|{\cal F}_{m-1}\} , \  \ m=\overline{1,\infty}, \ \  j=\overline{1,n}.$$
Let us put
\begin{eqnarray}\label{kj5}
 \bar \Psi^{j}_m=g_m - g_{m-1} - E^{P_j}\{g_m - g_{m-1}|{\cal F}_{m-1}\}.
\end{eqnarray}
The assumptions of  Theorem \ref{kj1} and  Lemma \ref{l3},  the representation (\ref{kj5}) imply 
$E^{P_j}|\bar \Psi^{j}_m|< 4(f_0+T), $  $E^{P_j}\{\bar \Psi^{j}_m|{\cal F}_{m-1}\}=0, \ j=\overline{1,n}, \ m=\overline{1,\infty}. $  
This proves  the necessity.

 {\bf The sufficiency.}  For any set of measures $P_1, \ldots,P_n \in M$ the representation (\ref{kj2}) for a non-decreasing adapted  process  $g=\{g_m\}_{m=0}^{\infty}, \ g_0=0,$   is valid. Hence, we obtain (\ref{kj4})    and (\ref{kj3}). The  equalities (\ref{kj4}), (\ref{kj3}) and the formula
\begin{eqnarray*}\label{nsal}
 E^{P}\{f_{m}+g_ m|{\cal F}_{m-1}\}=\frac{\sum\limits_{i=1}^{n}\alpha_i E^{P_1}\{\varphi_i|{\cal F}_{m-1}\}E^{P_i}\{f_{m}+g_m|{\cal F}_{m-1}\}}{\sum\limits_{i=1}^{n}\alpha_i E^{P_1}\{\varphi_i|{\cal F}_{m-1}\}}, \quad P \in M_n,
\end{eqnarray*}
$$\varphi_i=\frac{dP_i}{dP_1}, \quad i=\overline{1,n}, $$
imply
$$ E^{P}\{f_m+g_m|{\cal F}_{m-1}\}=f_{m-1}+g_{m-1}, \quad  m=\overline{1,\infty}, \quad  P \in M_n.$$ 
Arbitrariness of the set of measures $P_1, \ldots,P_n \in M$  and  fulfilment of the condition (\ref{buti1}) for the supermartingale $\{f_m, {\cal F}_m\}_{m=0}^\infty$  imply its   regularity.

Further, we consider a class of supermartingales $ F$ satisfying conditions

\begin{eqnarray*}\label{f1}
\sup\limits_{P\in M}E^P|f_m|< \infty, \quad m=\overline{0,\infty}. 
\end{eqnarray*}
\begin{defin} A  supermartingale $f=\{f_{m},\ {\cal F}_m\}_{m=0}^{\infty}  \in F$  is said to be    local  regular one if there exists an increasing  sequence of  nonrandom stopping times $\tau_{k_s} = k_s, \  k_s < \infty, \  s=\overline{1, \infty}, \  \lim\limits_{s \to \infty}k_s=\infty,  $ such that the stopped process  $f^{\tau_{k_s}}=\{f_{m\wedge \tau_{k_s}},\ {\cal F}_m\}_{m=0}^{\infty}$
 is a regular supermartingale for every $\tau_{k_s} = k_s, \  k_s < \infty, \  s=\overline{1, \infty}.$
\end{defin}
\begin{thm}\label{hf1}
Let $\{f_m, {\cal F}_m\}_{m=0}^\infty$ be a supermartingale relative to  a convex set of equivalent measures $M,$ belonging to the class $F,$
for which the representation
\begin{eqnarray}\label{jk1}
 f_m=M_m -g_m^0,  \quad m=\overline{0,\infty},
\end{eqnarray} 
 is valid, where $\{M_m\}_{m=0}^\infty$ is a martingale  relative to a convex set of equivalent measures $M$ such that
$$E^P|M_m|< \infty, \quad m=\overline{0,\infty}, \quad P \in M,  $$
 $g^0=\{g_m^0\}_{m=0}^\infty, \  g_0^0=0,$ is a non-decreasing adapted process. Then $\{f_m, {\cal F}_m\}_{m=0}^\infty$ is a  local regular supermartingale.
\end{thm}
{\bf Proof.}
The representation (\ref{jk1})  and assumptions of  Theorem \ref{hf1} imply inequalities
$ E^Pg_m^0<\infty, \ m=\overline{1,\infty}, \ P \in M.$
For any  measure $P \in M,$ therefore we have
\begin{eqnarray}\label{jk2}
 E^{P}\{f_m+g_m^0|{\cal F}_{m-1}\}=M_{m-1}=f_{m-1}+g_{m-1}^0, \quad  m=\overline{1,\infty}.
\end{eqnarray}
Consider a sequence of stopping times $\tau_s=s, \ s=\overline{1, \infty}.$
Equalities (\ref{jk2}) yield
\begin{eqnarray}\label{mif1}
 E^{P}\{f_{m\wedge\tau_s} +g_{m\wedge\tau_s}^0|{\cal F}_{m-1}\}=M_{(m-1)\wedge\tau_s}=
f_{(m-1)\wedge\tau_s}+g_{(m-1)\wedge\tau_s}^0,
\end{eqnarray}
$$  m=\overline{1,\infty}, \quad P \in M. $$
For the stopped supermartingale   $\{f_{m\wedge\tau_s}, {\cal F}_m\}_{m=0}^\infty,$
the set $G$ of adapted non-decreasing processes $g=\{g_m\}_{m=0}^\infty, \  g_0=0,$  such that   $\{f_{m\wedge\tau_s}+g_m, {\cal F}_m\}_{m=0}^\infty$
 is a supermartingale  relative to a convex set of equivalent measures $M$
contains nonzero element $g^{0, \tau_s}=$  $\{g_{m\wedge\tau_s}^0\}_{m=0}^\infty,$ $  g_0^0=0.$
Consider a maximal chain $\tilde G \subseteq G$  containing   this element  and let $g=\{g_m\}_{m=0}^\infty, \  g_0=0,$ be a maximal element in $\tilde G$ which exists, since the stopped supermartingale   $\{f_{m\wedge\tau_s}, {\cal F}_m\}_{m=0}^\infty$ is such that $|f_{m\wedge\tau_s}| \leq \sum\limits_{i=0}^s|f_i|=\varphi, \  m=\overline{0,\infty}, \  E^P\varphi \leq \sum\limits_{i=0}^s\sup\limits_{P \in M}E^P|f_i|=T< \infty.$
 Then 
\begin{eqnarray}\label{jk3}
 E^{P}\{f_{m\wedge\tau_s}+g_m|{\cal F}_{m-1}\}\leq f_{(m-1)\wedge\tau_s}+g_{m-1}, \quad  m=\overline{1,\infty}.
\end{eqnarray}
Equalities (\ref{mif1}) and inequality $ g^{0, \tau_s} \preceq g$ imply
\begin{eqnarray}\label{0jk4}
f_0 =  E^{P}\{f_{m\wedge\tau_s}+g_{m\wedge\tau_s}^0\}\leq E^{P}\{f_{m\wedge\tau_s}+g_m\} \leq f_0, \  m=\overline{1,\infty}, \ P \in M.
\end{eqnarray}
The last  inequalities yield 
\begin{eqnarray}\label{jk4}
 E^{P}\{f_{m\wedge\tau_s}+g_m\} = f_0, \quad  m=\overline{1,\infty}, \quad  P \in M.
\end{eqnarray}
 The equalities (\ref{jk4}), inequality  $  g^{0, \tau_s} \preceq g,$ and equalities
\begin{eqnarray}\label{jk5}
 E^{P}\{f_{m\wedge\tau_s}+g_{m\wedge\tau_s}^0\}=M_{0}=f_{0}, \quad  m=\overline{1,\infty}, \quad  P \in M,
\end{eqnarray}
imply that $ g^{0, \tau_s}=g.$   

So, we proved that the stopped supermartingale
 $\{f_{m\wedge\tau_s}, {\cal F}_m\}_{m=0}^\infty$ is regular one for every stopping time $\tau_s, \  s=\overline{1, \infty},$  converging to  the infinity,
as $ s \to \infty.$ This proves Theorem \ref{hf1}.

\begin{thm}\label{mars1}  On  a measurable space $\{\Omega, {\cal F}\},$
let a supermartingale $\{ f_m, {\cal F}_m\}_{m=0}^\infty $ relative to  a convex set of equivalent measures $M$ belongs to a class $F$   and there exists a nonnegative adapted  random process $\{\bar  g_m^0\}_{m=1}^\infty, \ E^P \bar g_m^0< \infty, \  m=\overline{1, \infty}, \  P \in M,$ such that 
\begin{eqnarray}\label{mars2}
f_{m-1} - E^P\{f_m|{\cal F}_{m-1}\}=E^P\{\bar g_m^0|{\cal F}_{m-1}\}, \quad m=\overline{1, \infty}, \quad P\in M,
\end{eqnarray}
then $\{ f_m, {\cal F}_m\}_{m=0}^\infty $ is a local regular supermartingale.
\end{thm}
{\bf Proof.} To prove Theorem \ref{mars1} let us consider a random process
$$\bar M_m=f_m+\sum\limits_{i=1}^m \bar g_i^0, \quad m=\overline{1, \infty}, \quad P\in M, \quad  f_0=\bar M_0.$$
 It is evident that
$E^P|\bar M_m | < \infty, \ m=\overline{1,\infty}, \ P \in M,$ and $E^P\{\bar M_m|{\cal F}_{m-1}\}=\bar M_{m-1}, \ m=\overline{1, \infty}, \ P\in M.$
Therefore, for $f_m$ the representation 
\begin{eqnarray}\label{mars3}
f_{m} =\bar M_m - g_{m}, \quad m=\overline{0, \infty}, 
\end{eqnarray}
is valid, where  $g_m=\sum\limits_{i=1}^m\bar g_i^0.$
Supermartingale (\ref{mars3}) satisfies conditions of the Theorem 6.
The Theorem  \ref{mars1} is proved.

\section{Description of local regular supermartingales.}

Below, we describe local regular supermartingales. For this we need some auxiliary statements.

Let $P_1, \ldots, P_k $ be a  family of  equivalent  measures on a measurable space   $\{ \Omega, {\cal F}\}$ and let us introduce denotation $M$ for a convex set of equivalent measures
 $$M=\left\{Q, \  Q=\sum\limits_{i=1}^{k}\alpha_i P_i, \ \alpha_i \geq 0, \  i=\overline{1,k},\ \sum\limits_{i=1}^{k}\alpha_i=1\right\}.$$

\begin{lemma}\label{q2}
If  $\xi$  is an  integrable random value relative to the set of equivalent  measures $P_1, \ldots, P_k $, then the formula
\begin{eqnarray}\label{n2}
\mathrm{ess}\sup\limits_{Q \in M}E^{Q}\{\xi|{\cal F}_n\}=\max\limits_{1 \leq i \leq k}E^{P_i}\{\xi|{\cal F}_n \}
\end{eqnarray}
is valid almost everywhere relative to the  measure $P_1$.
\end{lemma}
{\bf Proof.} The definition for $\mathrm{ess}\sup$ of non countable family of random variable see \cite{Chow}. Using the formula
\begin{eqnarray}\label{n3}
 E^{Q}\{\xi|{\cal F}_n\}=\frac{\sum\limits_{i=1}^{k}\alpha_i E^{P_1}\{\varphi_i|{\cal F}_n\}E^{P_i}\{\xi|{\cal F}_n\}}{\sum\limits_{i=1}^{k}\alpha_i E^{P_1}\{\varphi_i|{\cal F}_n\}}, \quad Q \in M,
\end{eqnarray}
 where $\varphi_i=\frac{dP_i}{dP_1},$ we obtain the inequality 
$$ E^{Q}\{\xi|{\cal F}_n\} \leq \max\limits_{1 \leq i \leq k}E^{P_i}\{\xi|{\cal F}_n\}, \quad Q \in M,$$  
or,
$$\mathrm{ess}\sup\limits_{Q \in M} E^{Q}\{\xi|{\cal F}_n\} \leq \max\limits_{1 \leq i \leq k}E^{P_i}\{\xi|{\cal F}_n\}.$$
On the other side  \cite{Chow},  
$$ E^{P_i}\{\xi|{\cal F}_n\} \leq \mathrm{ess}\sup\limits_{Q \in M} E^{Q}\{\xi|{\cal F}_n\}, \quad i=\overline{1,k}.$$ 
Therefore,
$$\max\limits_{1 \leq i \leq k} E^{P_i}\{\xi|{\cal F}_n\} \leq \mathrm{ess}\sup\limits_{Q \in M} E^{Q}\{\xi|{\cal F}_n\} .$$ 
The Lemma \ref{q2} is proved.

\begin{lemma}\label{hj1}
Let $G$ be a sub $\sigma$-algebra of  $\sigma$-algebra ${\cal F}$ and $f_s, s \in S,$ be a nonnegative bounded family of  random values relative to every measure from $M.$ Then 
\begin{eqnarray}\label{rgps1}
 E^P\{\mathrm{ess}\sup\limits_{s \in S}f_s|G\}\geq \mathrm{ess}\sup\limits_{s \in S}E^P\{f_s|G\}, \quad P \in M. \end{eqnarray}
\end{lemma}
{\bf Proof.} From the definition of $\mathrm{ess}\sup$ \cite{Chow}, we have the inequalities
\begin{eqnarray}\label{gps1}
\mathrm{ess}\sup\limits_{s \in S} f_s \geq f_t, \quad t \in S.
\end{eqnarray}
Therefore,
\begin{eqnarray}\label{gps2}
E^P\{\mathrm{ess}\sup f_s|G\}\geq  E^P\{ f_t|G\}, \quad t \in S.
\end{eqnarray}
The last implies
\begin{eqnarray}\label{gps3}
E^P\{\mathrm{ess}\sup\limits_{s \in S} f_s|G\}\geq \mathrm{ess}\sup\limits_{s \in S} E^P\{f_s|G\}.
\end{eqnarray}

In the next Lemma, we present formula for calculation of conditional expectation relative to another measure from $M.$
\begin{lemma}\label{q1}
 On a measurable space  $\{ \Omega, {\cal F}\}$ with filtration ${\cal F}_n$ on it, 
let $M$ be a convex  set of equivalent measures  and let  $\xi$ be a bounded random value.  Then the following formulas 
\begin{eqnarray}\label{n1}
E^{P_1}\{\xi|{\cal F}_n\}=E^{P_2}\left\{\xi \varphi_n^{P_1}|{\cal F}_n\right\}, \quad n=\overline{1, \infty},   
\end{eqnarray}
are valid,  where
\begin{eqnarray*}
 \varphi_n^{P_1}=\frac{dP_1}{dP_2}\left[E^{P_2}\left\{\frac{dP_1}{dP_2}|{\cal F}_n\right\}\right]^{-1}, \quad P_1, \ P_2 \in M.
\end{eqnarray*}
\end{lemma}
{\bf Proof.} The proof of the Lemma \ref{q1}   is evident.

\begin{lemma}\label{q3} On  a measurable space $\{ \Omega, {\cal F}\}$ with filtration ${\cal F}_n$ on it,  let $\xi$ be a nonnegative bounded random value. 
 Then the formulas 
\begin{eqnarray}\label{0n3}
 E^{Q}\{\mathrm{ess}\sup\limits_{P\in M} E^{P}\{\xi|{\cal F}_n\}|{\cal F}_m\}=
 \mathrm{ess}\sup\limits_{P\in M}E^{Q}\{\xi  \varphi_n^{P}|{\cal F}_m\}, \quad Q \in M, \quad n>m,  
\end{eqnarray}
are valid, where
$$ \varphi_n^{P}=\frac{dP}{dQ}\left[E^{Q}\left\{\frac{dP}{dQ}|{\cal F}_n\right\}\right]^{-1}, \quad P \in M.$$
\end{lemma}
{\bf Proof.}  From  the Lemma \ref{q1}, we obtain
$$\mathrm{ess}\sup\limits_{P\in M}E^{P}\{\xi|{\cal F}_n\}= \mathrm{ess}\sup\limits_{P\in M}E^{Q}\{\xi  \varphi_n^{P}|{\cal F}_n\}, \quad Q \in M.   $$
Due to Lemma \ref{hj1}, we obtain the inequality
$$E^{Q}\{\mathrm{ess}\sup\limits_{P\in M}E^{P}\{\xi|{\cal F}_n\}|{\cal F}_m\}=
E^{Q}\{\mathrm{ess}\sup\limits_{P\in M}E^{Q}\{\xi  \varphi_n^{P}|{\cal F}_n\}|{\cal F}_m\} \geq $$
$$\mathrm{ess}\sup\limits_{P\in M}E^{Q}\{\xi  \varphi_n^{P}|{\cal F}_m\}.$$
Let us prove reciprocal inequality
$$E^{Q}\{\mathrm{ess}\sup\limits_{P\in M}E^{P}\{\xi|{\cal F}_n\}|{\cal F}_m\}\leq
\mathrm{ess}\sup\limits_{P\in M}E^{Q}\{\xi  \varphi_n^{P}|{\cal F}_m\}.$$
From the definition of 
$\mathrm{ess}\sup\limits_{P\in M}E^{Q}\{\xi  \varphi_n^{P}|{\cal F}_n\},$
there exists a countable set $D=\{\bar P_i \in M, i=\overline{1, \infty}\}$  \cite{Chow}  such that
$$ \mathrm{ess}\sup\limits_{P\in M}E^{Q}\{\xi  \varphi_n^{P}|{\cal F}_n\}  =\sup\limits_{P\in D}E^{Q}\{\xi  \varphi_n^{P}|{\cal F}_n\}.  $$
The sequence $\varphi_k=\sup\limits_{P\in D}E^{Q}\{\xi  \varphi_n^{P}|{\cal F}_n\} - \max\limits_{1 \leq i \leq k} E^{Q}\{\xi  \varphi_n^{\bar P_i}|{\cal F}_n\}, \  k=\overline{1, \infty},$
converges to zero with probability one, as $k$  tends to infinity. 
  It is evident that 
$$ \max\limits_{1 \leq i \leq k}E^{Q}\{\xi  \varphi_n^{\bar P_i}|{\cal F}_n\}=E^{Q}\{ \xi \varphi_n^{\bar P_{\tau_k}}|{\cal F}_n\},$$
where
$$\tau_1=1,$$
$$ \tau_i=\left\{\begin{array}{l l} \tau_{i-1},  &   E^{Q}\{ \xi \varphi_n^{\bar P_{\tau_{i-1}}}|{\cal F}_n\} > E^{Q}\{  \xi \varphi_n^{\bar P_i} |{\cal F}_n\}, \\ 
i, &  E^{Q}\{ \xi \varphi_n^{\bar P_i}|{\cal F}_n\} \geq E^{Q}\{\xi \varphi_n^{\bar P_{\tau_{i-1}}}|{\cal F}_n\},
\end{array} \right. \quad i=\overline{2,\infty}.
$$
Therefore,
$$E^{Q}\{\mathrm{ess}\sup\limits_{P\in M}E^{P}\{\xi|{\cal F}_n\}|{\cal F}_m\}=
E^{Q}\{\mathrm{ess}\sup\limits_{P\in M}E^{Q}\{\xi  \varphi_n^{P}|{\cal F}_n\}|{\cal F}_m\} =$$
$$=E^{Q}\{\sup\limits_{P\in D}E^{Q}\{\xi  \varphi_n^{P}|{\cal F}_n\}|{\cal F}_m\}=
E^{Q}\{\lim\limits_{k \to \infty}\max\limits_{1 \leq i \leq k}E^{Q}\{\xi  \varphi_n^{\bar P_{i}}|{\cal F}_n\}|{\cal F}_m\}=$$
$$\lim\limits_{k \to \infty} E^{Q}\{E^{Q}\{\xi  \max\limits_{1 \leq i \leq k} \varphi_n^{\bar P_{i}}|{\cal F}_n\}|{\cal F}_m\}=
\lim\limits_{k \to \infty} E^{Q}\{\xi \varphi_n^{\bar P_{\tau_k}}|{\cal F}_m\}\leq$$
$$\mathrm{ess}\sup\limits_{P\in M} E^{Q}\{\xi  \varphi_n^{ P}|{\cal F}_m\}.$$
 In equalities above, we can  change the limits under conditional expectation sign,  since  with probability one the inequalities 
$$ \max\limits_{1 \leq i \leq k} \varphi_n^{\bar P_{i}} \leq   \max\limits_{1 \leq i \leq k+1} \varphi_n^{\bar P_{i}}, \quad k=\overline{1,\infty},$$
are valid.
Lemma \ref{q3} is proved.

The next Lemma is proved, as Lemma \ref{q3}.
\begin{lemma}\label{qq3} 
On a measurable space $\{ \Omega, {\cal F}\}$ with filtration ${\cal F}_n$ on it,
 let $\xi$ be a nonnegative bounded random value. 
Then the equalities
\begin{eqnarray}\label{nn3}
 E^{Q}\{\xi \mathrm{ess}\sup\limits_{P\in M}    \varphi_n^{P}|{\cal F}_n\}=
\mathrm{ess}\sup\limits_{P\in M}  E^{Q}\{\xi  \varphi_n^{P}|{\cal F}_n\}, \quad Q \in M, \quad n=\overline{0,\infty}, 
\end{eqnarray}
are valid, where
$$ \varphi_n^{P}=\frac{dP}{dQ}\left[E^{Q}\left\{\frac{dP}{dQ}|{\cal F}_n\right\}\right]^{-1}.$$
\end{lemma}

\begin{lemma}\label{lkq4} 
 On a measurable space $\{ \Omega, {\cal F}\}$ with filtration ${\cal F}_n$ on it,  for every nonnegative bounded random value $\xi$ the inequalities 
\begin{eqnarray}\label{lkn4}
 E^{Q}\{\mathrm{ess}\sup\limits_{P\in M} E^{P}\{\xi |{\cal F}_n\}|{\cal F}_m\}\leq
\mathrm{ess}\sup\limits_{P\in M} E^{P}\{\xi {\cal F}_m\}, \quad n>m, 
\end{eqnarray}
are  valid.
\end{lemma}
{\bf Proof.}
From the  Lemma \ref{q3}, we have
$$ E^{Q}\{\sup\limits_{P\in D} E^{P}\{\xi|{\cal F}_n\}|{\cal F}_m\}=
E^{Q}\{\sup\limits_{P\in D}E^{Q}\{\xi  \varphi_n^{P}|{\cal F}_n\}|{\cal F}_m\}=$$
\begin{eqnarray}\label{nnj30}
\sup\limits_{P\in D}E^{Q}\{\xi  \varphi_n^{P}|{\cal F}_m\}, \quad n>m,  
\end{eqnarray}
where $D$  is a countable subset of the set   $ M.$ Without loss of generality, we assume that the set of measures $\{P_1, \ldots, P_k\}$ belongs to the countable set $D=\{\bar P_i \in M, i=\overline{1, \infty}\}.$ First, we assume that $Q \in D.$ 
Then, it is evident that the following equalities 
\begin{eqnarray}\label{por1}
 \bigcup\limits_{i=1}^\infty\left\{\omega, \ E^{Q}\left\{\frac{d \bar P_i}{dQ}|{\cal F}_n\right\}\geq E^{Q}\left\{\frac{ d \bar P_i}{dQ}|{\cal F}_m\right\}\right\} =\Omega,  \ n > m, 
\end{eqnarray}
are valid.
Due to  (\ref{por1}), for every $\omega \in \Omega$ there exist  $1 \leq i < \infty $ such that
\begin{eqnarray}\label{por3}
\frac{\xi \frac{d \bar P_i}{dQ}}{E^Q\{\frac{d \bar P_i}{dQ}|{\cal F}_n\}} \leq \frac{\xi \frac{d\bar P_i}{dQ}}{E^Q\{\frac{d\bar P_i}{dQ}|{\cal F}_m\}}.
\end{eqnarray}
Therefore,
\begin{eqnarray}\label{ppor4}
\sup\limits_{\bar P_i \in D}\frac{\xi \frac{d \bar P_i}{dQ}}{E^Q\{\frac{d \bar P_i}{dQ}|{\cal F}_n\}} \leq \sup\limits_{\bar P_i \in D} \frac{\xi \frac{d\bar P_i}{dQ}}{E^Q\{\frac{d\bar P_i}{dQ}|{\cal F}_m\}}.
\end{eqnarray}
From (\ref{ppor4}), we obtain the  inequality
\begin{eqnarray}\label{por4}
E^{Q}\{\sup\limits_{\bar P_i \in D}\frac{\xi \frac{d \bar P_i}{dQ}}{E^Q\{\frac{d \bar P_i}{dQ}|{\cal F}_n\}}|{\cal F}_m\} \leq
 E^{Q}\{\sup\limits_{\bar P_i \in D} \frac{\xi \frac{d\bar P_i}{dQ}}{E^Q\{\frac{d\bar P_i}{dQ}|{\cal F}_m\}}|{\cal F}_m\}.
\end{eqnarray}
Or,
\begin{eqnarray}\label{port4}
E^{Q}\{E^{Q}\{\sup\limits_{\bar P_i \in D}\frac{\xi \frac{d \bar P_i}{dQ}}{E^Q\{\frac{d \bar P_i}{dQ}|{\cal F}_n\}}|{\cal F}_n\}|{\cal F}_m\} \leq
 E^{Q}\{\sup\limits_{\bar P_i \in D} \frac{\xi \frac{d\bar P_i}{dQ}}{E^Q\{\frac{d\bar P_i}{dQ}|{\cal F}_m\}}|{\cal F}_m\}.
\end{eqnarray}
The Lemmas  \ref{q3},  \ref{qq3} and  inequality (\ref{port4}) prove  Lemma \ref{lkq4}, as $Q \in D.$  Let  $Q  \in M.$  Since the set of measures $\{P_1, \ldots, P_k\}$ belongs to $D$ we complete the proof of  the Lemma  \ref{lkq4},     using the formula
\begin{eqnarray}\label{port5}
 E^{Q}\{\Phi|{\cal F}_{m}\}=\frac{\sum\limits_{i=1}^{k}\alpha_i E^{P_1}\{\varphi_i|{\cal F}_{m}\}E^{P_i}\{\Phi|{\cal F}_{m}\}}{\sum\limits_{i=1}^{k}\alpha_i E^{P_1}\{\varphi_i|{\cal F}_{m}\}}, \quad Q \in M,
\end{eqnarray}
and proved above inequalities, as $Q \in D,$  where $\Phi=\sup\limits_{\bar P_i \in D}E^{\bar P_i}\{\xi|{\cal F}_n\},  \varphi_i=\frac{dP_i}{dP_1}, \ i=\overline{1,k}.$
 The Lemma  \ref{lkq4} is proved.

\begin{lemma}\label{q5} On a measurable space $\{ \Omega, {\cal F}\}$ with filtration ${\cal F}_n$ on it,
let  $\xi$ be an  integrable  relative to the set of  equivalent measures $P_1, \ldots, P_k$  random value. 
Then the inequalities
\begin{eqnarray}\label{start6}
 E^Q\{\max\limits_{1 \leq i \leq k} E^{P_i}\{\xi|{\cal F}_n\}|{\cal F}_m\} \leq 
\max\limits_{1 \leq i \leq k} E^{P_i}\{\xi|{\cal F}_m\}, \quad n>m, \quad Q \in M,
\end{eqnarray}
are valid.
\end{lemma}
{\bf Proof.} Using the Lemma \ref{q2} and the Lemma  \ref{lkq4} for a bounded
$\xi,$ we prove the Lemma \ref{q5} inequalities (\ref{start6}).
Let us consider the case, as 
$\max\limits_{1 \leq i \leq k}  E^{P_i}\xi < \infty.$ Let $\xi_s, s=\overline{1, \infty},$ be a sequence of bounded random values  converging to $\xi$ monotonuosly. Then 
\begin{eqnarray}\label{alkn44}
 E^{Q}\{\max\limits_{1 \leq i \leq k} E^{P_i}\{\xi_s|{\cal F}_n\}|{\cal F}_m\}\leq
\max\limits_{1 \leq i \leq k}  E^{Q}\{\xi_s |{\cal F}_m\}, \quad l=\overline{1,k}.
\end{eqnarray}
Due to monotony convergence  of $\xi_s$ to $\xi,$ as $s \to \infty,$ we can pass to the limit under conditional expectations on the left and on the right in inequalities (\ref{alkn44}) that proves the  Lemma \ref{q5}.

\begin{lemma}\label{1q5} 
On a measurable space $\{ \Omega, {\cal F}\}$ with filtration ${\cal F}_m$ on it, 
let  $\xi$ be a nonnegative integrable random value with respect to a set of equivalent measures $\{P_1, \ldots, P_k \}$  and such that 
\begin{eqnarray}\label{r7}
 E^{P_i}\xi=M_0, \quad i=\overline{1,k},
\end{eqnarray}
then the random process  $\{M_m=\mathrm{ess}\sup\limits_{P\in M}E^P\{\xi|{\cal F}_m\}, {\cal F}_m\}_{m=0}^\infty$ is a martingale relative to a convex set of equivalent measures  $M.$
\end{lemma}
{\bf Proof.} Due to Lemma  \ref{q5}, a random process $\{M_m=\mathrm{ess}\sup\limits_{P\in M}E^P\{\xi|{\cal F}_m\}, {\cal F}_m\}_{m=0}^\infty $  is a supermartingale, that is, 
$$ E^P\{M_m | {\cal F}_{m-1}\} \leq M_{m-1}, \quad m=\overline{1, \infty}, \quad P \in M.$$
Or, $E^PM_m \leq M_0.$
From the other side,
$$ E^{P_s} [ \max\limits_{1\leq i \leq k}E^{P_i}\{\xi|{\cal F}_m\}] \ge 
\max\limits_{1\leq i \leq k}E^{P_s}E^{P_i}\{\xi|{\cal F}_m\} \geq M_0, \quad s=\overline{1,k}.$$
 The above inequalities imply $E^{P_s}M_m= M_0, \  m=\overline{1, \infty}, \ s=\overline{1, k}.$
The last equalities lead to equalities $E^{P}M_m= M_0, \ m=\overline{1, \infty}, \ P \in M.$
The fact that $M_m$ is a supermartingale relative to the set of measures $M$ and the above equalities  prove the  Lemma \ref{1q5}, since the Lemma \ref{l1} conditions are valid.

\begin{thm}\label{mars12}
On a measurable space $\{ \Omega, {\cal F}\}$ with filtration ${\cal F}_m$ on it,  let  $\xi$ be a  ${\cal F}_N$-measurable nonnegative integrable   relative to a set of equivalent measures $\{P_1, \ldots, P_k \}$ random value, $N < \infty.$  Then
 a supermartingale $\{f_m, {\cal F}_m\}_{m=0}^\infty, $ where 
\begin{eqnarray}\label{marsss13}
f_m=\mathrm{ess}\sup\limits_{P \in M}E^P\{\xi| {\cal F}_m\}, \quad  m=\overline{1,\infty}, \quad   \max\limits_{1 \leq i \leq k}E^{P_i}\xi< \infty, 
\end{eqnarray}
 is  local regular one  if and only if
\begin{eqnarray}\label{mars13}
E^{P_i}\xi=f_0, \quad i=\overline{1,k}.
\end{eqnarray}
\end{thm}
{\bf Proof.} { \bf The necessity.} Let $\{f_m, {\cal F}_m\}_{m=0}^\infty $  be a local regular supermartingale. Then there exists a sequence of nonrandom stopping times $\tau_s=n_s, \ s=\overline{1, \infty},$ such that for every $n_s$ there exists
$ \varphi=\sum\limits_{m=1}^{n_s}\sum\limits_{i=1}^{k}E^{P_i}\{\xi|{\cal F}_m\}$ satisfying inequalities
$$\max\limits_{1 \leq j \leq k}E^{P_j}\varphi \leq \sum\limits_{m=1}^{n_s}\sum\limits_{i=1}^{k}\max\limits_{1 \leq j \leq k}E^{P_j}
E^{P_i}\{\xi|{\cal F}_m \}\leq$$
$$\sum\limits_{m=1}^{n_s}\sum\limits_{i=1}^{k}\max\limits_{1 \leq j \leq k}E^{P_j} \max\limits_{1 \leq i \leq k} E^{P_i}\{\xi|{\cal F}_m\} \leq$$
$$ \sum\limits_{m=1}^{n_s}\sum\limits_{i=1}^{k}\max\limits_{1 \leq j \leq k}E^{P_j} \max\limits_{1 \leq i \leq k} E^{P_i} \xi =n_s k  \max\limits_{1 \leq i \leq k} E^{P_i} \xi,$$
$$\sup\limits_{P \in M}E^P\varphi \leq  \max\limits_{1 \leq j \leq k}E^{P_j}\varphi\leq n_s k  \max\limits_{1 \leq i \leq k} E^{P_i} \xi,$$
and nonnegative adapted   random process $\{\bar g_m^0\}_{m=0}^\infty, \ \bar g_0^0=0, \ $
$E^{P_i}\bar g_m^0< \infty, \ 0 \leq m \leq n_s$ such that 
$$ f_m +\sum\limits_{i=1}^m\bar g_i^0=\bar M_m, \quad E^{P}\bar M_m=f_0, \quad 0 \leq m \leq n_s, \quad P \in M.$$ 
If $n_s >N,$ then 
$$E^{P_i}(\xi+\sum\limits_{i=1}^N \bar g_i^0)=E^{P_i}\xi+ E^{P_i}\sum\limits_{i=1}^N\bar g_i^0=f_0.$$
But there exists $1 \leq  i_1 \leq k$ such that $E^{P_{i_1}}\xi=f_0.$ Therefore,
$E^{P_{i_1}}\sum\limits_{i=1}^Ng_i^0=0.$ Due to equivalence of measures $P_i, \ i=\overline{1,k},$ we obtain 
\begin{eqnarray}\label{mars14}
E^{P_i}\xi=f_0, \quad i=\overline{1,k},
\end{eqnarray}
 where $f_0=\sup\limits_{P \in M}E^P\xi.$

{\bf Sufficiency.} If conditions  (\ref{mars14}) are satisfied,  then  $\{\bar M_m, {\cal F}_m\}_{m=0}^\infty $ is a martingale, where  $\bar M_m= \sup\limits_{P \in M}E^P\{\xi|{\cal F}_m\}.$ The last implies local regularity of $\{f_m, {\cal F}_m\}_{m=0}^\infty.$
The Theorem \ref{mars12} is proved.

Below we consider an arbitrary convex set of equivalent measures $M$ on a measurable space $\{\Omega, {\cal F}\}$ and a filtration ${\cal F}_n$ on it.  Introduce into consideration a set  $A_0$ of all integrable  nonnegative random values $\xi$  relative to a convex set of equivalent measures $M$ satisfying conditions
\begin{eqnarray}\label{0mars6}
E^P\xi=1, \quad P \in M.
\end{eqnarray}
It is evident that the set  $A_0$ is not empty, since contains random value $\xi =1.$
More interesting case is  as  $A_0$ contains more then one element.

\begin{lemma}\label{tmars5}  On measurable space $\{\Omega, {\cal F}\}$ and a filtration ${\cal F}_n$ on it, let $M$ be an arbitrary convex set of equivalent 
measures. If  non negative random value $\xi$ is such that $\sup\limits_{P \in M}E^P\xi< \infty,$  then 
$\{f_m=\mathrm{ess}\sup\limits_{P \in M}E^P\{\xi| {\cal F}_m\},  {\cal F}_m\}_{m=0}^\infty $ is a supermartingale relative to the convex set of equivalent measures $M.$
\end{lemma}
{\bf Proof.} From definition of $\mathrm{ess}\sup$ \cite{Chow}, for every $\mathrm{ess}\sup\limits_{P \in M}E^P\{\xi| {\cal F}_m\}$ there exists a countable set $D_m$ such that 
$$\mathrm{ess}\sup\limits_{P \in M}E^P\{\xi| {\cal F}_m\}=\sup\limits_{P \in D_m}E^P\{\xi| {\cal F}_m\}, \quad m=\overline{0, \infty}.$$ 
 The set $D=\bigcup\limits_{m=0}^\infty D_m$ is also countable and
\begin{eqnarray}\label{tamar1}  
\mathrm{ess}\sup\limits_{P \in M}E^P\{\xi| {\cal F}_m\}=\sup\limits_{P \in D}E^P\{\xi| {\cal F}_m\}.
\end{eqnarray} 
Really, since
\begin{eqnarray}\label{tamar2}
 \sup\limits_{P \in D}E^P\{\xi| {\cal F}_m\} \geq \sup\limits_{P \in D_m}E^P\{\xi| {\cal F}_m\}=\mathrm{ess}\sup\limits_{P \in M}E^P\{\xi| {\cal F}_m\}.
\end{eqnarray}
From the other side,
\begin{eqnarray}\label{tamar3}
\mathrm{ess}\sup\limits_{P \in M}E^P\{\xi| {\cal F}_m\} \geq E^Q\{\xi| {\cal F}_m\}, \quad Q \in M.
\end{eqnarray}
The last gives 
\begin{eqnarray}\label{tamar4}
\mathrm{ess}\sup\limits_{P \in M}E^P\{\xi| {\cal F}_m\} \geq \sup\limits_{P \in D}E^P\{\xi| {\cal F}_m\}. 
\end{eqnarray}
The inequalities (\ref{tamar2}),  (\ref{tamar4}) prove the needed. So, for all $m$ we can choose the common set $D.$ Let $D=\{\bar P_1,\ldots \bar P_n, \ldots\}.$ Due to Lemma  \ref{q5}, for every $Q \in \bar M_k,$ where
\begin{eqnarray}\label{tamar5}
\bar M_k=\{P \in M, P=\sum\limits_{i=1}^k\alpha_i \bar P_i, \  \alpha_i \geq 0, \ \sum\limits_{i=1}^k\alpha_i=1\},
\end{eqnarray}
we have
\begin{eqnarray}\label{tamar6}
 E^Q\{\max\limits_{1 \leq i \leq k} E^{\bar P_i}\{\xi|{\cal F}_n\}|{\cal F}_m\} \leq 
\max\limits_{1 \leq i \leq k} E^{\bar P_i}\{\xi|{\cal F}_m\}, \quad n>m, \quad Q \in \bar M_k,
\end{eqnarray}
It is evident that $\max\limits_{1 \leq i \leq k} E^{\bar P_i}\{\xi|{\cal F}_n\}$ tends to $ \sup\limits_{P \in D}E^P\{\xi| {\cal F}_n\}$ monotonously increasing, as $k \to \infty.$ Fixing $Q \in \bar M_k \subset \bar M_{k+1}$ and tending $k $ to the infinity in 
inequalities (\ref{tamar6}) we obtain
\begin{eqnarray}\label{tamar7}
 E^Q\{\sup\limits_{P \in D}E^P\{\xi| {\cal F}_n\}| {\cal F}_m\} \leq 
\sup\limits_{P \in D}E^P\{\xi| {\cal F}_m\}, \quad n>m, \quad Q \in \bar M_k,
\end{eqnarray}
The last inequalities implies that for every measure $Q,$ belonging to the convex span, constructed on the set $D,$ $\{f_m=\mathrm{ess}\sup\limits_{P \in M}E^P\{\xi| {\cal F}_m\},  {\cal F}_m\}_{m=0}^\infty $ is a supermartingale relative to the convex set of equivalent measures, generated by set  $D.$ Now, if a measure $Q_0$ does not  belong to the convex span, constructed on the set $D,$ then we can  add it to the set $D$ and repeat the proof made above. As a result, we proved that  $\{f_m=\mathrm{ess}\sup\limits_{P \in M}E^P\{\xi| {\cal F}_m\},  {\cal F}_m\}_{m=0}^\infty $ is also a supermartingale relative to the measure $Q_0.$ The Zorn Lemma \cite{Kelley} complete the proof of the Lemma \ref{tmars5}.

\begin{thm}\label{fmars5} On measurable space $\{\Omega, {\cal F}\}$ and a filtration ${\cal F}_n$ on it, let $M$ be an arbitrary convex set of equivalent 
measures.  For a random value $\xi \in A_0$  the random process $\{ E^P\{\xi|{\cal F}_m\}, {\cal F}_m\}_{m=0}^\infty,$ $P \in M,$  is a local regular martingale  relative to a convex set of equivalent 
measures $M.$ 
\end{thm}
{\bf Proof.}  Let $P_1, \ldots, P_n$ 
be a certain subset of measures from $M.$ Denote by
$M_n$  a convex set of equivalent measures 
\begin{eqnarray}\label{mars8}
M_n = \{P \in M, \  P=\sum\limits_{i=1}^n\alpha_i P_i, \  \alpha_i \geq 0, \ i=\overline{1,n}, \ \sum\limits_{i=1}^n\alpha_i=1\}.
\end{eqnarray}
Due to Lemma \ref{1q5}, $\{\bar M_m, {\cal F}_m\}_{m=0}^\infty$ is a  martingale relative
to the set of measures $M_n,$ where $ \bar M_m=\mathrm{ess}\sup\limits_{P \in M_n}E^P\{\xi|{\cal F}_m\}, \ \xi \in A_0.$  Let us consider an arbitrary measure $P_0 \in M$ and let
\begin{eqnarray}\label{mars9}
M_n^{P_0} = \{P \in M, \ P= \sum\limits_{i=0}^n\alpha_i P_i, \  \alpha_i \geq 0, \  i=\overline{0,n}, \ \sum\limits_{i=0}^n\alpha_i=1\}.
\end{eqnarray}
Then  $\{\bar M_m^{P_0}, {\cal F}_m\}_{m=0}^\infty,$ where $\bar M_m^{P_0}=\mathrm{ess}\sup\limits_{P \in M_n^{P_0}}E^P\{\xi|{\cal F}_m\},$ is a martingale relative to the set of measures $M_n^{P_0}.$ It is evident that
\begin{eqnarray}\label{mars10}
\bar M_m \leq \bar M_m^{P_0}, \quad   m=\overline{0, \infty}.
\end{eqnarray}
Since $E^P\bar M_m=E^P\bar M_m^{P_0}=1, \ m=\overline{0, \infty}, \ P \in M_n,$ the inequalities (\ref{mars10}) give $\bar M_m=\bar M_m^{P_0}.$ Analogously,
$E^{P_0}\{\xi|{\cal F}_m\} \leq \bar M_m^{P_0}.$ From equalities 
$ E^{P_0}E^{P_0}\{\xi|{\cal F}_m\}$ $ = E^{P_0}\bar M_m^{P_0}=1$ we obtain
$E^{P_0}\{\xi|{\cal F}_m\} = \bar M_m^{P_0}=\bar M_m.$ 
Since the measure $P_0$ is arbitrary it implies that $\{E^P\{\xi|{\cal F}_m\}, {\cal F}_m\}_{m=0}^\infty$ is a martingale relative to all measures from $M.$
Due to Theorem  \ref{mars1}, it is a local regular supermartingale with random process $\bar g^0_m=0,  m=\overline{0, \infty}. $
The Theorem \ref{fmars5} is proved.

\begin{thm}\label{mmars1} On measurable space $\{\Omega, {\cal F}\}$ and a filtration ${\cal F}_n$ on it, let $M$ be an arbitrary convex set of equivalent 
measures. 
If $\{f_m, {\cal F}_m\}_{m=0}^\infty$ is an adapted random process  satisfying conditions
\begin{eqnarray}\label{mmars2}
f_m \leq f_{m-1}, \quad  E^P\xi|f_m| <\infty, \quad P \in M \quad m=\overline{1, \infty}, \quad  \xi \in A_0,
\end{eqnarray}
then the random process
\begin{eqnarray}\label{mmars3}
 \{ f_mE^P\{\xi|{\cal F}_m\}, {\cal F}_m\}_{m=0}^\infty, \quad  P \in M,
\end{eqnarray}
is a local regular supermartingale relative to a convex set of equivalent  measures  $M.$
\end{thm}

{\bf Proof.} Due to Theorem  \ref{fmars5}, the random process 
$\{ E^P\{\xi|{\cal F}_m\}, {\cal F}_m\}_{m=0}^\infty$ is a martingale relative to the convex set of equivalent  measures  $M.$ Therefore,
\begin{eqnarray*}
f_{m-1}E^P\{\xi|{\cal F}_{m-1}\} - E^P\{ f_m E^P\{\xi|{\cal F}_m\}|{\cal F}_{m-1}\}=
\end{eqnarray*}
\begin{eqnarray}\label{mmars4}
E^P\{ (f_{m-1} - f_m) E^P\{\xi|{\cal F}_m\}|{\cal F}_{m-1}\}, \quad m=\overline{1, \infty}.
\end{eqnarray}
So, if to put  $\bar g_m^0= (f_{m-1} - f_m) E^P\{\xi|{\cal F}_m\}, \ m=\overline{1, \infty}, $ then $\bar g_m^0 \geq 0,$  it is  ${\cal F}_m$-measurable and
$E^P\bar g_m^0 \leq E^P\xi(|f_{m-1}|+|f_m|)< \infty.$ It proves the needed statement.

\begin{cor}\label{hg1} If $f_m=\alpha, \ m=\overline{1, \infty}, \ \alpha \in R^1,$ $\xi \in A_0,$ then 
$\{\alpha E^P\{\xi |{\cal F}_m\},  {\cal F}_m \}_{m=0}^\infty$ is a local regular martingale. Assume that  $\xi =1,$ then $\{f_m, {\cal F}_m\}_{m=0}^\infty$ is a local regular supermartingale relative to a convex set of equivalent  measures  $M.$ 
\end{cor}
Denote by $F_0$ the set of adapted processes
\begin{eqnarray*}\label{mmars5}
F_0=\{ f=\{f_m\}_{m=0}^\infty,  \   P(|f_m| <\infty) =1, \ P \in M, \ f_m \leq f_{m-1}, \  m=\overline{1, \infty}\}.
\end{eqnarray*}
For every $\xi \in A_0$ let us introduce the set of adapted processes
$$ L_{\xi}=$$
\begin{eqnarray*}\label{mmars6}
\{\bar f=\{f_mE^P\{\xi|{\cal F}_m\}\}_{m=0}^\infty, \  \{f_m\}_{m=0}^\infty \in F_0, \   E^P\xi|f_m| <\infty, \ P \in M, \ m=\overline{1, \infty}\},
\end{eqnarray*}
and 
\begin{eqnarray*}\label{mmars7}
V=\bigcup\limits_{\xi \in A_0}L_{\xi}.
\end{eqnarray*}

\begin{cor}\label{fdr1} Every  random  process from the set $K,$ where
\begin{eqnarray}\label{mmars88}
K=\left \{ \sum\limits_{i=1}^mC_i \bar f_i, \ \bar f_i \in V, \  C_i \geq 0, \ i=\overline{1, m}, \ m=\overline{1, \infty}\right\}, 
\end{eqnarray}
 is a local regular supermartingale relative to the convex set of equivalent  measures  $M$  on a measurable space $\{\Omega, {\cal F}\}$ with filtration ${\cal F}_m$ on it. 
\end{cor}
{\bf Proof.} The proof is evident.

\begin{thm}\label{mmars9} On measurable space $\{\Omega, {\cal F}\}$ and a filtration ${\cal F}_n$ on it, let $M$ be an arbitrary convex set of equivalent 
measures. 
 Suppose that  $\{f_m, {\cal F}_m\}_{m=0}^\infty$ is a nonnegative uniformly integrable supermartingale relative to a convex set of equivalent  measures  $M,$ then 
the necessary and sufficient conditions for it  to be a local regular one is belonging it to the set $K.$
\end{thm}
{\bf Proof.}
{\bf Necessity.}  It is evident  that if  $\{f_m, {\cal F}_m\}_{m=0}^\infty$ belongs to $K,$ then it is a local regular supermartingale.

{\bf Sufficiency.} Suppose that  $\{f_m,{\cal F}_m\}_{m=0}^\infty$ is a local regular supermartingale. Then there exists  nonnegative adapted process $\{\bar g_m^0\}_{m=1}^ \infty,  \ E^P\bar g_m^0< \infty, \ m=\overline{1, \infty}, $  and a martingale  $\{M_m\}_{m=0}^ \infty,$
such that 
\begin{eqnarray*}\label{mmars8}
f_m=M_m - \sum\limits_{i=1}^m\bar g_i^0, \quad  m=\overline{0, \infty}. 
\end{eqnarray*}
Then $M_m \geq 0, \ m=\overline{0, \infty}, \ E^P M_m <\infty, \ P\in M.$
Since $0< E^PM_m=f_0< \infty$ we have $E^P\sum\limits_{i=1}^m\bar g_i^0< f_0.$ Let us put $g_{\infty}=\lim\limits_{m \to \infty}\sum\limits_{i=1}^m\bar g_i^0.$
Using uniform integrability of $f_m,$ we can pass to the limit in the equality
$$ E^P(f_m +\sum\limits_{i=1}^m\bar g_i^0)=f_0, \quad P \in M,$$
as $m \to \infty$.
Passing to the limit in the last equality, as $m \to \infty,$  we obtain
$$E^P(f_\infty +g_{\infty})=f_0, \quad P \in M.$$
Introduce into consideration a  random value $\xi=\frac{f_\infty +g_{\infty}}{f_0}.$
Then $E^P\xi=1, \ P \in M.$   From here we obtain that  $\xi \in A_0$ and 
$$M_m=f_0E^P\{\xi|{\cal F}_m\}, \ m=\overline{0, \infty}.$$
 Let us put $\bar f_m^2=-\sum\limits_{i=1}^m\bar g_i^0. $ It is easy to see that an adapted random process $\bar f_2=\{\bar f_m^2, {\cal F}_m\}_{m=0}^\infty$ belongs to $F_0.$ Therefore,
for the supermartingale $f=\{ f_m,{\cal F}_m\}_{m=0}^\infty$ the representation 
$$f=\bar f_1+ \bar f_2,$$
is valid, where $\bar f_1=\{f_0E^P\{\xi|{\cal F}_m\}, {\cal F}_m\}_{m=0}^ \infty$  belongs to $L_{\xi}$
with  $ \xi = \frac{f_\infty +g_{\infty}}{f_0}$  and $ f_m^1=f_0, \ m=\overline{0,\infty}.$ The same is valid for $\bar f_2$ with $\xi=1.$ This implies that $f$ belongs to the set $K.$ The Theorem 
\ref{mmars9} is proved.

\begin{cor}\label{mars16}  Let $f_N, \  N< \infty,$ be a ${\cal F}_N$-measurable integrable random value,  $\sup\limits_{P \in M} E^P|f_N| < \infty,$ and let there exist $\alpha_0 \in R^1$ such that
$$ -\alpha_0  M_N+ f_N \leq 0, \quad \omega \in \Omega, $$
where $\{ M_m, {\cal F}_m\}_{m=0}^\infty=\{E^P\{\xi|{\cal F}_m\}, {\cal F}_m\}_{m=0}^\infty, \ \xi \in A_0. $ 
Then a  supermartingale $\{ f_m^0+ \bar f_m\}_{m=0}^\infty$ is local regular one relative to  a convex set of equivalent measures $M,$ where
$$f_m^0=\alpha_0 M_m,  $$
$$\bar f_m=
\left\{\begin{array}{l l} 0, & m<N, \\
f_N - \alpha_0 M_N, & m \geq N.  
 \end{array} \right. $$
\end{cor}
{\bf Proof.}  It is evident that $\bar f_{m-1} -\bar f_m \geq 0, \  m=\overline{0,\infty}.$
Therefore, the supermartingale
$$f_m^0+ \bar f_m=
\left\{\begin{array}{l l} \alpha_0 M_m, & m<N, \\
f_N , & m= N,  \\
f_N - \alpha_0 M_N+\alpha_0 M_m, & m>N,
 \end{array} \right. $$
is local regular one relative to  a convex set of equivalent measures $M.$
The Corollary \ref{mars16} is proved.

\section{Optional decomposition for non negative supermartingales.} 
 In this section we introduce the notion of complete set of equivalent measures and prove that  non negative supermartingales are local regular with respect to this set of measures. For this purpose  we are needed the next auxiliary statement.
\begin{thm}\label{nick1}
The necessary and sufficient condition of local regularity of non negative supermartingale $\{f_m, {\cal F}_m\}_{m=0}^\infty$ relative to a convex set of equivalent measures $M$ is the existence of ${\cal F}_m$-measurable random value $\xi_m^0 \in A_0$ such that
\begin{eqnarray}\label{nick2}
\frac{f_m}{f_{m-1}} \leq \xi_m^0, \quad E^P\{\xi_m^0|{\cal F}_{m-1}\}=1, \quad P\in M, \quad m=\overline{1, \infty}.
\end{eqnarray}
\end{thm}
{\bf Proof.} {\bf The necessity.} Let  $\{f_m, {\cal F}_m\}_{m=0}^\infty$ be a local regular supermartingale. Then there exists non negative adapted random process
$\{g_m\}_{m=0}^\infty, \ g_0=0,$ such that $\sup\limits_{P \in M}E^Pg_m<\infty,$
\begin{eqnarray}\label{nick3}
f_{m-1} - E^P\{f_m|{\cal F}_{m-1}\} =E^P\{g_m|{\cal F}_{m-1}\}, \quad P \in M, \quad m=\overline{1, \infty}.
\end{eqnarray}
Let us put $\xi_m^0=\frac{f_m+g_m}{f_{m-1}}, \ m=\overline{1, \infty}.$ Then
from (\ref{nick3}) $E^P\{\xi_m^0|{\cal F}_{m-1}\}=1, \ P \in M, \ m=\overline{1, \infty}.$
It is evident that inequalities (\ref{nick2}) are valid.

{\bf The sufficiency.} Suppose that conditions of the Theorem \ref{nick1} are valid.
Then   $ f_m \leq f_{m-1}+ f_{m-1}(\xi_m^0 -1).$
Introduce  denotation  $g_m= -f_m+ f_{m-1}\xi_m^0.$ Then  $g_m \geq 0, $  $\sup\limits_{P \in M}E^Pg_m \leq \sup\limits_{P \in M}E^Pf_m + \sup\limits_{P \in M}E^Pf_{m-1}<\infty, \ m=\overline{1, \infty}.$  The last inequalities and equality   give
\begin{eqnarray}\label{nick4}
f_m=f_0+\sum\limits_{i=1}^m f_{i-1}(\xi_i^0 -1) - \sum\limits_{i=1}^m g_{i}, \quad \ m=\overline{1, \infty}.
\end{eqnarray}
Let us consider  $\{M_m, {\cal F}_m\}_{m=0}^\infty,$ where $M_m=f_0+\sum\limits_{i=1}^m f_{i-1}(\xi_i^0 -1).$ Then  $E^P\{M_m| {\cal F}_{m-1}\}$ $=M_{m-1}, \ P\in M, \ m=\overline{1, \infty}.$
The Theorem \ref{nick1} is proved.

\subsection{Space of  finite set of elementary events.}

In this subsection we assume that  a space of elementary events  $\Omega$ is finite, that is,  $N_0=|\Omega|< \infty,$  and we give new proof of optional decomposition for non negative supermartingale relative to some convex set of equivalent measures.

 Let  ${\cal F}$ be   some algebra of subsets of    $\Omega$  and let  ${\cal F}_n \subset {\cal F}_{n+1} \subset {\cal F} $ be an increasing set of  algebras, where   ${\cal F}_0 =\{\emptyset, \Omega\}, $
  ${\cal F}_N = {\cal F}. $ Denote by $M$ a set of equivalent measures on a measurable space $\{\Omega, {\cal F}\}.$ Further, we assume that a set $A_0$ contains  an element $\xi_0\neq 1.$  It is evident that every algebra ${\cal F}_n$ is generated by sets $A_i^n, \   i=\overline{1, N_n}, A_i^n\cap A_j^n=\emptyset, \ i \neq j, \ N_n<\infty, \ \bigcup\limits_{i=1}^{N_n}A_i^n=\Omega, \ n=\overline{1,N}.$
Between the sets $A_i^n$ and $A_j^{n-1}$ the relations $A_j^{n-1}=\bigcup\limits_{s \in I_j}A_s^n$  are valid, where $I_j \subseteq T_n, \ T_n=\{1,2,\ldots,N_n\}, I_s\cap I_k=\emptyset, \ s\neq k, \  \bigcup\limits_{j=1}^{N_{n-1}}I_j=T_n.$
Let  $m_n=E^P\{\xi_0|{\cal F}_n\}, \ P \in M,  \   n=\overline{1, N}.$ Then for $m_n$ the representation 
\begin{eqnarray}\label{1myk1}
m_n=\sum\limits_{i=1}^{N_n}m_i^n\chi_{A_i^n}(\omega), \quad n=\overline{1,N},
\end{eqnarray} 
is valid.
Consider the difference $m_n -m_{n-1}.$ Then

$$m_n -m_{n-1}=\sum\limits_{s=1}^{N_{n-1}}\sum\limits_{j \in I_s}(m_j^n -m_s^{n-1})\chi_{A_j^n}(\omega)=$$
\begin{eqnarray}\label{1myk2}
\sum\limits_{s=1}^{N_{n-1}}\sum\limits_{j=1}^{N_n}\chi_{I_s}(j)(m_j^n - m_s^{n-1})\chi_{A_j^n}=
 \sum\limits_{j=1}^{N_{n}}[m_j^n - \sum\limits_{s=1}^{N_{n-1}}\chi_{I_s}(j)m_s^{n-1}]\chi_{A_j^n}.
\end{eqnarray}
Introduce the  set of  numbers $a_{js}^n=m_j^n - m_s^{n-1}, j \in I_s, \ s=\overline{1,N_{n-1}},$ 
and  sets $I_s^-=\{j \in I_s, \ a_{js}^n \leq 0\}, $ \  $  I_s^+=\{j \in I_s, \ a_{js}^n > 0\},$ \  $I^-=\bigcup\limits_{s=1}^{N{n-1}}I_s^-,$ \ $I^+=\bigcup\limits_{s=1}^{N{n-1}}I_s^+.$
Then 
\begin{eqnarray}\label{1myk3}
m_n -m_{n-1}=\sum\limits_{j\in I^-}d_{j}^n\chi_{A_j^n}(\omega)+\sum\limits_{j\in I^+}d_{j}^n\chi_{A_j^n}(\omega), 
\end{eqnarray}
\begin{eqnarray}\label{1myk4}
\sum\limits_{j\in I^-}\chi_{A_j^n}(\omega)+
\sum\limits_{j\in I^+}\chi_{A_j^n}(\omega)=1, 
\end{eqnarray}
where $d_j^n=a_{js}^n,$ as $ j \in I_s^-,$ or $j \in I_s^+.$
From equalities (\ref{1myk3}), (\ref{1myk4})   we obtain
\begin{eqnarray}\label{1myk5}
\sum\limits_{j\in I^-}d_{j}^nP(A_j^n)+\sum\limits_{j\in I^+}d_{j}^nP(A_j^n)=0, \quad P  \in M,
\end{eqnarray}
\begin{eqnarray}\label{1myk6}
\sum\limits_{j\in I^-}P(A_j^n)+
\sum\limits_{j\in I^+}P(A_j^n)=1, \quad  \in M.
\end{eqnarray}
Denote by $M_n$ the contraction of the set of measures $M$ on the algebra ${\cal F}_n.$  Introduce into the set $M_n $ metrics
\begin{eqnarray}\label{1myk7} 
\rho_n(P_1,P_2)=\sum\limits_{j \in I^-}|P_1(A_j^n) - P_2(A_j^n)|+
\end{eqnarray}
$$ \sum\limits_{j \in I^+}|P_1(A_j^n) - P_2(A_j^n)|, \quad n=\overline{1,N}.$$

\begin{defin}\label{1myk8} On a measurable space
$ \{\Omega, {\cal F}\},$  a set of measure $M$ we call complete if for every $1 \leq n \leq N$
the closure of the set of measures $M_n$ in metrics (\ref{1myk7}) contains  measures
\begin{eqnarray}\label{1myk9} 
P_{ij}^n(A)= \left\{\begin{array}{l l} 0, & A\neq A_i^n, A_j^n,\\
                                                    \frac{d_j^n}{-d_i^n +d_j^n}, & A=A_i^n,\\
                                                      \frac{-d_i^n}{-d_i^n +d_j^n}, & A=A_j^n,
                                                     \end{array}\right.       
\end{eqnarray}
for every  $i \in I^-$ and $ j \in I^+.$
\end{defin}

\begin{lemma}\label{1myk10} Let a family of measures $M$ be complete and the set $A_0$ contains an element $\xi_0\neq 1.$ Then for every non negative ${\cal F}_n$-measurable random value $\xi_n=\sum\limits_{i=1}^{N_n} C_i^n \chi_{A_i^n}$ there exists a real number $\alpha_n$ such that
\begin{eqnarray}\label{1myk11} 
\frac{ \sum\limits_{i=1}^{N_n} C_i^n \chi_{A_i^n} }{\sup\limits_{P \in M_n}\sum\limits_{i=1}^{N_n} C_i^n P(A_i^n)} \leq 1+\alpha_n (m_n - m_{n-1}), \quad 
n=\overline{1,N}.
\end{eqnarray}
\end{lemma}
{\bf Proof.} 
On the set  $\bar M_n,$  a  functional $\varphi(P)=\sum\limits_{i=1}^{N_n} C_i^n P(A_i^n)$  is continuous one, where  $ \bar M_n$ is the closure of the set $M_n$ in the metrics $\rho_n(P_1,P_2).$
From this it follows that  the equality
\begin{eqnarray}\label{1myk13}
\sup\limits_{P \in M_n}\sum\limits_{i=1}^{N_n} C_i^n P(A_i^n)=\sup\limits_{P \in \bar M_n}\sum\limits_{i=1}^{N_n} C_i^n P(A_i^n)
\end{eqnarray} 
is valid.
 Denote by $f_i^n=\frac{C_i^n}{\sup\limits_{P \in M_n}\sum\limits_{i=1}^{N_n} C_i^n P(A_i^n)}, \ i=\overline{1, N_n}.$ Then
\begin{eqnarray}\label{1myk14}
\sum\limits_{i=1}^{N_n} f_i^n P(A_i^n) \leq 1, \quad P \in \bar M_n.
\end{eqnarray} 
In every set  $I_s^-$ there are strictly negative elements and  in the every  set $ I_s^+$  there are strictly positive elements. For those $i \in I^-$ for which $d_i^n<0$ and those $j \in I^+$ for which $d_j^n>0$  the inequality (\ref{1myk14}) is as follows 
\begin{eqnarray}\label{1myk15}
\quad  f_i^n \frac{d_j^n}{-d_i^n +d_j^n} + \frac{-d_i^n}{-d_i^n+d_j^n}f_j^n\leq 1,
\end{eqnarray} 
$$   d_i^n <0,  \quad  i \in I^-,  \quad d_j ^n>0,  \quad  j \in I^+. $$
From  (\ref{1myk15})  we obtain inequalities
\begin{eqnarray}\label{1myk16}
f_j^n \leq 1+\frac{1- f_i^n}{-d_i^n}d_j^n, \quad d_i^n <0,  \quad i \in I^-,  \quad d_j^n >0,  \quad  j \in I^+.
\end{eqnarray} 
Since the inequalities  (\ref{1myk16}) are valid for every $\frac{1- f_i^n}{-d_i^n},$ as $ d_i^n <0, $ and since the set of such  elements is finite,  then if to denote
$$ \alpha_n =\min_{\{i, \ d_i^n <0\}}\frac{1- f_i^n}{-d_i^n},$$
 then we have
\begin{eqnarray}\label{1myk17}
f_j^n \leq 1+\alpha_n d_j^n,   \quad  d_j^n >0,  \quad  j \in I^+.
\end{eqnarray} 
From the definition of  $\alpha_n$ we obtain inequalities 
\begin{eqnarray*}
f_i^n \leq 1+\alpha_n d_i^n,  \quad   d_i^n <0, \quad  i \in I^-.
\end{eqnarray*} 
Now if $d_i^n=0$ for some $ i \in I^-,$ then in this case $f_i^n \leq 1.$ All these inequalities give 
\begin{eqnarray}\label{1myk18}
f_i^n \leq 1+\alpha_n d_i^n,  \quad   i \in I^-\cup I^+.
\end{eqnarray} 
   Multiplying on $\chi_{A_i^n}$ the inequalities (\ref{1myk18}) and summing over all $ i \in I^-\cup I^+$ we obtain the needed inequality.
The Lemma  \ref{1myk10}    is proved.

\begin{thm}\label{1myk19} Suppose that conditions of the Lemma \ref{1myk10}
are valid. Then for every non negative supermartingale  $\{f_m, {\cal F}_m\}_{m=0}^N$ optional decomposition is valid.
\end{thm}
{\bf Proof.} Consider random value $\xi_n=\frac{f_n}{f_{n-1}}.$ Due to Lemma \ref{1myk10}
$$ \frac{\xi_n}{\sup\limits_{P \in M}E^P\xi_n} \leq 1+\alpha_n(m_n-m_{n-1})=\xi_n^0, \quad  n=\overline{1,N}.$$
It is evident that  $E^P\{\xi_n^0|{\cal F}_{n-1}\}=1, \ P \in M, \ n=\overline{1, N}.$ 
Since $\sup\limits_{P \in M} E^P\xi_n \leq 1,$ then 
\begin{eqnarray}\label{1myk20}
 \frac{f_n}{f_{n-1}} \leq \xi_n^0, \quad n=\overline{1, N}.
\end{eqnarray} 
The Theorem \ref{nick1} and inequalities (\ref{1myk20}) prove the Lemma \ref{1myk19}.

\subsection{Countable set of elementary events.}

In this subsection we generalize the results of the previous subsection onto the countable space of elementary events.

 Let  ${\cal F}$ be   some $\sigma$-algebra of subsets of  the countable set of elementary events  $\Omega$  and let  ${\cal F}_n \subset {\cal F}_{n+1} \subset {\cal F} $ be   a certain  increasing set  of $\sigma$-algebras, where   ${\cal F}_0 =\{\emptyset, \Omega\}.$
 Denote by $M$ a set of equivalent measures on a measurable space $\{\Omega, {\cal F}\}.$  Further, we assume that the set $A_0$ contains an element $\xi_0\neq 1.$  Suppose that   $\sigma$-algebra ${\cal F}_n$ is generated by sets $A_i^n, \   i=\overline{1, \infty}, \  A_i^n\cap A_j^n=\emptyset, \ i \neq j, \ \bigcup\limits_{i=1}^{\infty}A_i^n=\Omega, \ n=\overline{1,\infty}.$
We also assume that between the sets $A_i^n$ and $A_j^{n-1}$ the relations $A_j^{n-1}=\bigcup\limits_{s \in I_j}A_s^n$  are valid, where $I_j \subseteq N_0 =\{1,2,\ldots,n, \ldots\},$  $ I_s\cap I_k=\emptyset, \ s\neq k, \  \bigcup\limits_{j=1}^{\infty}I_j=N_0.$
Introduce into consideration a martingale  $m_n=E^P\{\xi_0|{\cal F}_n\}, \ P \in M, \  n=\overline{1, \infty}.$ Then for $m_n$ the representation 
\begin{eqnarray}\label{2myk1}
m_n=\sum\limits_{i=1}^{\infty}m_i^n\chi_{A_i^n}(\omega), \quad n=\overline{1,\infty},
\end{eqnarray} 
is valid.
Consider the difference $m_n -m_{n-1}.$ Then
$$m_n -m_{n-1}=\sum\limits_{s=1}^{\infty}\sum\limits_{j \in I_s}(m_j^n -m_s^{n-1})\chi_{A_j^n}(\omega)=$$
\begin{eqnarray}\label{2myk2}
\sum\limits_{s=1}^{\infty}\sum\limits_{j=1}^{\infty}\chi_{I_s}(j)(m_j^n - m_s^{n-1})\chi_{A_j^n}=
 \sum\limits_{j=1}^{\infty}[m_j^n - \sum\limits_{s=1}^{\infty}\chi_{I_s}(j)m_s^{n-1}]\chi_{A_j^n}.
\end{eqnarray}
Introduce the  set of  numbers $a_{js}^n=m_j^n - m_s^{n-1},  \  j \in I_s, \ s=\overline{1, \infty},$ 
and  sets $I_s^-=\{j \in I_s, \ a_{js}^n \leq 0\}, $ \  $  I_s^+=\{j \in I_s, \ a_{js}^n > 0\},$ \  $I^-=\bigcup\limits_{s=1}^{\infty}I_s^-,$ \ $I^+=\bigcup\limits_{s=1}^{\infty}I_s^+.$
Then 
\begin{eqnarray}\label{2myk3}
m_n -m_{n-1}=\sum\limits_{j\in I^-}d_{j}^n\chi_{A_j^n}(\omega)+\sum\limits_{j\in I^+}d_{j}^n\chi_{A_j^n}(\omega), 
\end{eqnarray}
\begin{eqnarray}\label{2myk4}
\sum\limits_{j\in I^-}\chi_{A_j^n}(\omega)+
\sum\limits_{j\in I^+}\chi_{A_j^n}(\omega)=1, 
\end{eqnarray}
where $d_j^n=a_{js}^n,$ as $ j \in I_s^-,$ or $j \in I_s^+.$
From equalities (\ref{2myk3}), (\ref{2myk4})   we obtain
\begin{eqnarray}\label{2myk5}
\sum\limits_{j\in I^-}d_{j}^nP(A_j^n)+\sum\limits_{j\in I^+}d_{j}^nP(A_j^n)=0, \quad P  \in M,
\end{eqnarray}
\begin{eqnarray}\label{2myk6}
\sum\limits_{j\in I^-}P(A_j^n)+
\sum\limits_{j\in I^+}P(A_j^n)=1, \quad P  \in M.
\end{eqnarray}
Denote by $M_n$ the contraction of the set of measures $M$ on the $\sigma$-algebra ${\cal F}_n.$  Introduce into the set $M_n $ metrics
\begin{eqnarray}\label{2myk7} 
\rho_n(P_1,P_2)=\sum\limits_{j \in I^-}|P_1(A_j^n) - P_2(A_j^n)|+\sum\limits_{j \in I^+}|P_1(A_j^n) - P_2(A_j^n)|,
\end{eqnarray}
$$  n=\overline{1, \infty}. $$
\begin{defin}\label{2myk8} On a measurable space
$ \{\Omega, {\cal F}\},$ a set of measure $M$  we call complete if for every $1 \leq n < \infty$
the closure of the set of measures $M_n$ in metrics (\ref{2myk7}) contains  measures
\begin{eqnarray}\label{2myk9} 
P_{ij}^n(A)= \left\{\begin{array}{l l} 0, & A\neq A_i^n, A_j^n,\\
                                                    \frac{d_j^n}{-d_i^n +d_j^n}, & A=A_i^n,\\
                                                      \frac{-d_i^n}{-d_i^n +d_j^n}, & A=A_j^n,
                                                     \end{array}\right.       
\end{eqnarray}
for every  $i \in I^-$ and $ j \in I^+.$
\end{defin}

\begin{lemma}\label{2myk10} Let a family of measures $M$ be complete and the set $A_0$ contains an element $\xi_0 \neq 1.$ Then for every non negative bounded ${\cal F}_n$-measurable random value $\xi_n=\sum\limits_{i=1}^{\infty} C_i^n \chi_{A_i^n}$ there exists real number $\alpha_n$ such that
\begin{eqnarray}\label{2myk11} 
\frac{ \sum\limits_{i=1}^{\infty} C_i^n \chi_{A_i^n} }{\sup\limits_{P \in M_n}\sum\limits_{i=1}^{\infty} C_i^n P(A_i^n)} \leq 1+\alpha_n (m_n - m_{n-1}), \quad 
n=\overline{1,\infty}.
\end{eqnarray}
\end{lemma}
{\bf Proof.}  
On the set  $\bar M_n,$ a functional $\varphi(P)=\sum\limits_{i=1}^{\infty} C_i^n P(A_i^n)$  is continuous  one, where  $ \bar M_n$ is the closure of the set $M_n$ in metrics $\rho_n(P_1,P_2).$
From this it follows that  the equality
\begin{eqnarray}\label{2myk13}
\sup\limits_{P \in M_n}\sum\limits_{i=1}^{\infty} C_i^n P(A_i^n)=\sup\limits_{P \in \bar M_n}\sum\limits_{i=1}^{\infty} C_i^n P(A_i^n)
\end{eqnarray} 
is valid.
Denote by $f_i^n=\frac{C_i^n}{\sup\limits_{P \in M_n}\sum\limits_{i=1}^{\infty} C_i^n P(A_i^n)}, \ i=\overline{1, \infty}.$
Then
\begin{eqnarray*}
\sum\limits_{i=1}^{\infty} f_i^n P(A_i^n) \leq 1, \quad P \in \bar M_n.
\end{eqnarray*}
The last inequalities can be written  in the form
\begin{eqnarray}\label{2myk14}
\sum\limits_{i \in I^-} f_i^n P(A_i^n) +\sum\limits_{i \in I^+}f_i^n P(A_i^n) \leq 1, \quad P \in \bar M_n.
\end{eqnarray}
 In every set  $I_s^-$ there are strictly negative elements and  in the every  set $ I_s^+$  there are strictly positive elements. For those $i \in I^-$ for which $d_i^n<0$ and those $j \in I^+$ for which $d_j^n>0$  the inequality (\ref{2myk14}) is as follows 
\begin{eqnarray}\label{2myk15}
 f_i^n \frac{d_j^n}{-d_i^n +d_j^n} + \frac{-d_i^n}{-d_i^n+d_j^n} f_j^n\leq 1, 
\end{eqnarray} 
$$ d_i^n <0, \quad d_j ^n>0, \quad  i \in I^-, \quad j \in I^+.$$
From  (\ref{2myk15})  we obtain inequalities
\begin{eqnarray}\label{2myk16}
f_j^n \leq 1+\frac{1- f_i^n}{-d_i^n}d_j^n, \quad d_i^n <0, \quad  d_j^n >0, 
\quad  i \in I^-, \quad j \in I^+.
\end{eqnarray} 
Two cases are possible: a) for all  $ i \in I^-,$ $ f_i^n \leq 1; $ b) there exists $ i \in I^-$ such that $  f_i^n > 1.$
First, let us consider the case a).

Since inequalities  (\ref{2myk16}) are valid for every $\frac{1- f_i^n}{-d_i^n},$ as $ d_i^n <0, $ and $ f_i^n \leq 1, i \in I^- ,$  then if to denote
$$ \alpha_n =\inf_{\{i, \ d_i^n <0\}}\frac{1- f_i^n}{-d_i^n},$$
 we have $ 0 \leq \alpha_n < \infty$  and
\begin{eqnarray}\label{2myk17}
f_j^n \leq 1+\alpha_n d_j^n,  \quad   d_j^n >0,  \quad j \in I^+.
\end{eqnarray} 
From the definition of  $\alpha_n$ we obtain inequalities 
\begin{eqnarray*}
f_i^n \leq 1+\alpha_n d_i^n,  \quad   d_i^n <0, \quad  i \in I^-.
\end{eqnarray*} 
Now, if $d_i^n=0$ for some $ i \in I^-, $ then in this case $f_i^n \leq 1.$ All these inequalities give 
\begin{eqnarray}\label{2myk18}
f_i^n \leq 1+\alpha_n d_i^n,  \quad   i \in I^-\cup I^+.
\end{eqnarray} 
Consider the case b). From the inequality (\ref{2myk16}) we obtain
\begin{eqnarray}\label{2myk19}
f_j^n \leq 1-\frac{1- f_i^n}{d_i^n}d_j^n, \quad d_i^n <0, \quad  d_j^n >0, \quad  i \in I^-, \quad j \in I^+.
\end{eqnarray} 
 The last inequalities give
\begin{eqnarray}\label{2myk20}
\frac{1- f_i^n}{d_i^n} \leq \min_{\{j, \ d_j^n >0\}} \frac{1}{d_j^n}< \infty, \quad  d_i^n <0, \quad  i \in I^-.
\end{eqnarray}
Let us define $\alpha_n =\sup\limits_{\{ i, \ d_i^n<0 \} }\frac{1- f_i^n}{d_i^n}< \infty.$
Then from (\ref{2myk19}) we obtain
\begin{eqnarray}\label{2myk21}
f_j^n \leq 1- \alpha_n d_j^n, \quad  \  d_j^n >0, \quad j \in I^+.
\end{eqnarray}
From the definition of  $\alpha_n $ we have
\begin{eqnarray}\label{2myk22}
f_i^n \leq 1- \alpha_n d_i^n, \quad  \  d_i^n <0, \quad  i \in I^-.
\end{eqnarray}
The inequalities (\ref{2myk21}), (\ref{2myk22}) give
\begin{eqnarray}\label{2myk23}
f_j^n \leq 1- \alpha_n d_j^n,  \quad  j \in I^-\cup I^+.
\end{eqnarray}

   Multiplying on $\chi_{A_j^n}$ the inequalities (\ref{2myk23}) and summing over all $ j \in I^-\cup I^+$ we obtain the needed inequality.
The Lemma  \ref{2myk10}  is proved.

\begin{thm}\label{2myk24} Suppose that conditions of the Lemma \ref{2myk10}
are valid. Then for every non negative supermartingale  $\{f_m, {\cal F}_m\}_{m=0}^\infty, $ satisfying conditions 
\begin{eqnarray}\label{2myk25}
\sup\limits_{P \in M}E^Pf_m < \infty, \quad  \frac{f_m}{f_{m-1}} \leq C_m< \infty, \quad m=\overline{1,\infty},
\end{eqnarray}
 optional decomposition is valid.
\end{thm}
{\bf Proof.} Consider random value $\xi_n=\frac{f_n}{f_{n-1}}.$ Due to Lemma \ref{2myk10}
$$ \frac{\xi_n}{\sup\limits_{P \in M}E^P\xi_n} \leq 1+\alpha_n(m_n-m_{n-1})=\xi_n^0.$$
It is evident that  $E^P\{\xi_n^0|{\cal F}_{n-1}\}=1, \ P \in M, \ n=\overline{1, \infty}.$ 
Since $\sup\limits_{P \in M}E^P\xi_n \leq 1,$ then 
\begin{eqnarray}\label{2myk26}
 \frac{f_n}{f_{n-1}} \leq \xi_n^0, \quad n=\overline{1, N}.
\end{eqnarray} 
The Theorem \ref{nick1} and inequalities (\ref{2myk26}) prove the Lemma \ref{2myk24}.

\subsection{An arbitrary space of elementary events.}

In this subsection we consider an arbitrary space of elementary events and prove optional decomposition for non negative supermartingales.

Let  ${\cal F}$ be   some $\sigma$-algebra of subsets of  the set of elementary events  $\Omega$  and let  ${\cal F}_n \subset {\cal F}_{n+1} \subset {\cal F} $ be an increasing set  of $\sigma$-algebras, where   ${\cal F}_0 =\{\emptyset, \Omega\}.$
 Denote by $M$ a set of equivalent measures on a measurable space $\{\Omega, {\cal F}\}.$ We assume that $\sigma$-algebras  ${\cal F}_n, \  n=\overline{1, \infty},$ and ${\cal F} $  are complete relative to all measure $P \in M.$   
Further, we suppose that a set $A_0$ contains an element $\xi_0\neq 1.$ 
Let  $m_n=E^P\{\xi_0|{\cal F}_n\}, \ P \in M, \ n=\overline{1, \infty}.$ Then for $m_n$ the representation 
\begin{eqnarray}\label{3myk1}
m_n=\sum\limits_{i=1}^{\infty}m_i^n\chi_{A_i^n}(\omega), \quad n=\overline{1,\infty},
\end{eqnarray} 
is valid for some $ A_i^n \in {\cal F}_n, $
$A_i^n, \   i=\overline{1, \infty}, \ A_i^n\cap A_j^n=\emptyset, \ i \neq j, \ \bigcup\limits_{i=1}^{\infty}A_i^n=\Omega, \ n=\overline{1,\infty}.$

Really, let us consider a sequence of random values  $m_n=E^P\{\xi_0|{\cal F}_n\}, \ P \in M, \  n=\overline{1,\infty}.$ It is evident that 
$E^P\{m_n|{\cal F}_{n-1}\}=m_{n-1}.$ For every random value  $m_n$ there exists not more then a countable set of non negative real number $m_s^n \geq 0$ such that $P(A_s^n)>0,$  where  $A_s^n=\{\omega \in \Omega, m_n=m_s^n\}.$  It is evident that  $A_i^n\cap A_j^n=\emptyset, \ i \neq j.$ Since $m_n$ is defined on all $\Omega,$ then $P(\bigcup\limits_{i=1}^{\infty}A_i^n)=1.$  From this it follows that 
$P(\Omega \setminus \bigcup\limits_{i=1}^{\infty}A_i^n)=0.$ The set $\Omega \setminus \bigcup\limits_{i=1}^{\infty}A_i^n$ we can  join, for example, to  $A_1^n$ and to put $m_n=m_1^n, \ \omega \in A_1^n\cup(\Omega \setminus \bigcup\limits_{i=1}^{\infty}A_i^n).$  
If to change denotation we  come to the above statement. Further, let us prove that we can choose the sets $A_i^n$ such that   the relations $A_j^{n-1}=\bigcup\limits_{s \in I_j}A_s^n$  are valid, where $I_j \subseteq N_0 =\{1,2,\ldots,n, \ldots\},$  $ I_s\cap I_k=\emptyset, \ s\neq k, \  \bigcup\limits_{j=1}^{\infty}I_j=N_0.$ Really, if it is not  so then we can choose countable set of subsets $B_{ij}^n=A_i^n\cap A_j^{n-1}, \ i, j =\overline{1, \infty}.$ It is evident that  $ \bigcup\limits_{j=1}^{\infty}B_{ij}^n=A_i^n,$
 $ \bigcup\limits_{i=1}^{\infty}B_{ij}^n= A_j^{n-1}.$
 For fixed $j$ denote by $I_j $ those indexes $i$ for which $P(B_{ij}^n)>0.$ Then
 $ \bigcup\limits_{i \in I_j}B_{ij}^n=A_j^{n-1}.$ Let us define on $ B_{ij}^n$ random value  putting $m_{ij}^n=m_i^n, \ \omega \in B_{ij}^n, \ i \in I_j, \ j=\overline{1, \infty}.$ Then 
\begin{eqnarray}\label{3myk2}
\sum\limits_{j=1}^{\infty}\sum\limits_{i \in I_j}m_{ij}^n\chi_{B_{ij}^n}(\omega)=
\sum\limits_{j=1}^{\infty}\sum\limits_{i \in I_j}m_{i}^n\chi_{B_{ij}^n}(\omega)=
m_n,  \quad n=\overline{1,\infty}.
\end{eqnarray} 

Taking into account these facts, further without loss of generality we  suppose that between the sets $A_i^n$ and $A_j^{n-1}$ the relations $A_j^{n-1}=\bigcup\limits_{s \in I_j}A_s^n$  are valid, where $I_j \subseteq N_0 =\{1,2,\ldots,n, \ldots\},$ \  $ I_s\cap I_k=\emptyset, \ s\neq k, \  \bigcup\limits_{j=1}^{\infty}I_j=N_0.$

 Consider the difference $m_n -m_{n-1}.$ Then
$$m_n -m_{n-1}=\sum\limits_{s=1}^{\infty}\sum\limits_{j \in I_s}(m_j^n -m_s^{n-1})\chi_{A_j^n}(\omega)=$$
\begin{eqnarray}\label{3myk2}
\sum\limits_{s=1}^{\infty}\sum\limits_{j=1}^{\infty}\chi_{I_s}(j)(m_j^n - m_s^{n-1})\chi_{A_j^n}=
 \sum\limits_{j=1}^{\infty}[m_j^n - \sum\limits_{s=1}^{\infty}\chi_{I_s}(j)m_s^{n-1}]\chi_{A_j^n}.
\end{eqnarray}
Introduce the  set of  numbers $a_{js}^n=m_j^n - m_s^{n-1}, j \in I_s, \ s=\overline{1, \infty},$ 
and  sets $I_s^-=\{j \in I_s, \ a_{js}^n \leq 0\}, $ \  $  I_s^+=\{j \in I_s, \ a_{js}^n > 0\},$ \  $I^-=\bigcup\limits_{s=1}^{\infty}I_s^-,$ \ $I^+=\bigcup\limits_{s=1}^{\infty}I_s^+.$
Then 
\begin{eqnarray}\label{3myk3}
m_n -m_{n-1}=\sum\limits_{j\in I^-}d_{j}^n\chi_{A_j^n}(\omega)+\sum\limits_{j\in I^+}d_{j}^n\chi_{A_j^n}(\omega), 
\end{eqnarray}
\begin{eqnarray}\label{3myk4}
\sum\limits_{j\in I^-}\chi_{A_j^n}(\omega)+
\sum\limits_{j\in I^+}\chi_{A_j^n}(\omega)=1, 
\end{eqnarray}
where $d_j^n=a_{js}^n,$ as $ j \in I_s^-,$ or $j \in I_s^+.$

Let a countable set of subsets  $ D_j^n \in {\cal F}_n, \ j=\overline{1, \infty},$  be such that $ D_i^n\cap D_j^n=\emptyset, \ i \neq j, \ \bigcup\limits_{i=1}^{\infty}D_i^n=\Omega, \ n=\overline{1,\infty}.$ Denote by
 $\tilde {\cal F}_n^{D} \subseteq {\cal F}_n$ a sub $\sigma$-algebra of the  $\sigma$-algebra ${\cal F}_n,$  generated by the countable set of subsets $ D_j^n \in {\cal F}_n, \ j=\overline{1, \infty}.$

Let  $M_n^D$ be the contraction of the set of measures $M$ on the   sub $\sigma$-algebra $\tilde {\cal F}_n^D \subseteq {\cal F}_n.$ 
 Introduce into the set $M_n^D $ metrics
\begin{eqnarray}\label{3myk7} 
\rho_n^D(P_1,P_2)=\sum\limits_{j \in I^-}|P_1(D_j^n) - P_2(D_j^n)|+\sum\limits_{j \in I^+}|P_1(D_j^n) - P_2(D_j^n)|, 
\end{eqnarray}
$$  n=\overline{1, \infty}. $$
\begin{defin}\label{3myk8}   On a measurable space
$ \{\Omega, {\cal F}\},$ a set of measure $M$ we call complete if for every $1 \leq n < \infty$ and every countable  set of subsets  $ D_j^n \in {\cal F}_n, \ j=\overline{1, \infty},$  $ D_i^n\cap D_j^n=\emptyset, \ i \neq j, \ \bigcup\limits_{i=1}^{\infty}D_i^n=\Omega, \ n=\overline{1,\infty},$  the closure   in metrics (\ref{3myk7}) of the  set of measures $M_n^D$ 
 contains  measures
\begin{eqnarray}\label{3myk9} 
P_{ij}^n(B)= \left\{\begin{array}{l l} 0, & B\neq D_i^n, D_j^n,\\
                                                    \frac{d_j^n}{-d_i^n +d_j^n}, & B=D_i^n,\\
                                                      \frac{-d_i^n}{-d_i^n +d_j^n}, & B=D_j^n,
                                                     \end{array}\right.       
\end{eqnarray}
for every  $i \in I^-$ and $ j \in I^+.$
\end{defin}

\begin{lemma}\label{3myk10} Let a family of measures $M$ be complete and the set $A_0$ contains an element $\xi_0 \neq 1.$ Then for every non negative bounded ${\cal F}_n$-measurable random value $\xi_n$ there exists a real number $\alpha_n$ such that
\begin{eqnarray}\label{3myk11} 
\frac{\xi_n }{\sup\limits_{P \in M}E^P\xi_n} \leq 1+\alpha_n (m_n - m_{n-1}), \quad 
n=\overline{1,\infty}.
\end{eqnarray}
\end{lemma}
{\bf Proof.}  For the random value $\xi_n$ the representation $\sum\limits_{j=1}^{\infty} \xi_j^n \chi_{V_j^n} $  is valid, where 
$ V_j^n \in {\cal F}_n, \ j=\overline{1, \infty},$  $ V_i^n\cap V_j^n=\emptyset, \ i \neq j, \ \bigcup\limits_{i=1}^{\infty}V_i^n=\Omega, \ n=\overline{1,\infty}.$
Introduce into consideration  the countable set of subsets 
$ U_{ij}^n=A_i^n\cap V_j^n, \ i, j =\overline{1, \infty}.$ It is evident that $\bigcup\limits_{i,j=1}^\infty U_{ij}^n=\Omega, $  \ $ U_{ij}^n \cap U_{rs}^n=\emptyset, \ \{i j\}\neq \{ r s \}. $

Then 
\begin{eqnarray}\label{33myk3}
m_n -m_{n-1}=\sum\limits_{s=1}^\infty\sum\limits_{i\in I^-}d_{i}^n\chi_{A_i^n \cap V_s^n}(\omega)+\sum\limits_{t=1}^\infty\sum\limits_{j\in I^+}d_{j}^n\chi_{A_j^n \cap V_t^n}(\omega), 
\end{eqnarray}
\begin{eqnarray}\label{33myk4}
\sum\limits_{s=1}^\infty\sum\limits_{i\in I^-}\chi_{A_i^n  \cap V_s^n}(\omega)+
\sum\limits_{t=1}^\infty\sum\limits_{j\in I^+}\chi_{A_j^n \cap V_t^n}(\omega)=1, 
\end{eqnarray}
where $d_j^n=a_{js}^n,$ as $ j \in I_s^-,$ or $j \in I_s^+.$
From equalities (\ref{33myk3}), (\ref{33myk4})   we obtain
\begin{eqnarray}\label{3myk5}
\sum\limits_{s=1}^\infty\sum\limits_{i\in I^-}d_{i}^nP(A_i^n  \cap V_s^n)+\sum\limits_{t=1}^\infty \sum\limits_{j\in I^+}d_{j}^nP(A_j^n \cap V_t^n)=0, \quad P  \in M,
\end{eqnarray}
\begin{eqnarray}\label{3myk6}
\sum\limits_{s=1}^\infty\sum\limits_{i\in I^-}P(A_i^n  \cap V_s^n)+
\sum\limits_{t=1}^\infty\sum\limits_{j\in I^+}P(A_j^n \cap V_t^n)=1, \quad P  \in M.
\end{eqnarray}
The random value $\xi_n$  can be written in the form
\begin{eqnarray}\label{33myk7}
\xi_n=\sum\limits_{j=1}^{\infty}\sum\limits_{s=1}^{\infty} \xi_s^n \chi_{A_j^n \cap V_s^n}.
\end{eqnarray}
Let $M_n^U$ be  the contraction of the set of measures $M$ on the sub $\sigma$-algebra $\tilde {\cal F}_n^U,$ generated by the countable set of subsets $U_{ij}^n, \ i,j=\overline{1, \infty}.$ 
On the set  $\bar M_n^U,$ a  functional $\varphi(P)=\sum\limits_{j=1}^{\infty}\sum\limits_{s=1}^{\infty} \xi_s^n P(A_j^n \cap V_s^n), \ P\in  \bar M_n^U,$  is continuous one in the metrics $\rho_n^U(P_1,P_2),$ where  $ \bar M_n^U$ is the closure of the set $M_n^U$ in the metrics.
From this it follows that  the equality
\begin{eqnarray}\label{3myk13}
\sup\limits_{P \in M_n^U}\sum\limits_{j=1}^{\infty}\sum\limits_{s=1}^{\infty} \xi_s^n P(A_j^n \cap V_s^n)=\sup\limits_{P \in \bar M_n^U}\sum\limits_{j=1}^{\infty}\sum\limits_{s=1}^{\infty} \xi_s^n P(A_j^n \cap V_s^n)
\end{eqnarray} 
is valid.

Denote by $f_s^n=\frac{\xi_s^n}{\sup\limits_{P \in M_n}\sum\limits_{j=1}^{\infty} \sum\limits_{s=1}^{\infty}\xi_s^n P(A_j^n\cap V_s^n)}, \ s=\overline{1, \infty}.$ Then
\begin{eqnarray}\label{3myk12}
\sum\limits_{j=1}^{\infty}\sum\limits_{s=1}^{\infty} f_s^n P(A_j^n \cap V_s^n) \leq 1, \quad P \in \bar M_n^U.
\end{eqnarray}
The last inequalities can be written  in the form
\begin{eqnarray}\label{33myk14}
\sum\limits_{s=1}^\infty\sum\limits_{i \in I^-} f_s^n P(A_i^n \cap V_s^n) +\sum\limits_{t=1}^\infty\sum\limits_{j \in I^+}f_t^n P(A_j^n  \cap V_t^n) \leq 1, \quad P \in \bar M_n^U.
\end{eqnarray}
Let us write equalities (\ref{33myk3}), (\ref{33myk4}) in  more general form 
\begin{eqnarray}\label{4myk3}
m_n -m_{n-1}=\sum\limits_{s=1}^\infty\sum\limits_{i\in I^-}d_{is}^n\chi_{A_i^n \cap V_s^n}(\omega)+\sum\limits_{t=1}^\infty\sum\limits_{j\in I^+}d_{jt}^n\chi_{A_j^n \cap V_t^n}(\omega), 
\end{eqnarray}
\begin{eqnarray}\label{4myk4}
\sum\limits_{s=1}^\infty\sum\limits_{i\in I^-}\chi_{A_i^n  \cap V_s^n}(\omega)+
\sum\limits_{t=1}^\infty\sum\limits_{j\in I^+}\chi_{A_j^n \cap V_t^n}(\omega)=1, 
\end{eqnarray}
where $d_{is}^n=d_{i}^n, \ s=\overline{1, \infty}, \  d_{jt}^n=d_{j}^n, \ t=\overline{1, \infty}.$ 

From equalities (\ref{4myk3}), (\ref{4myk4})   we obtain
\begin{eqnarray}\label{4myk5}
\sum\limits_{s=1}^\infty\sum\limits_{i\in I^-}d_{is}^nP(A_i^n  \cap V_s^n)+\sum\limits_{t=1}^\infty \sum\limits_{j\in I^+}d_{jt}^nP(A_j^n \cap V_t^n)=0, \quad P  \in M,
\end{eqnarray}
\begin{eqnarray}\label{4myk6}
\sum\limits_{s=1}^\infty\sum\limits_{i\in I^-}P(A_i^n  \cap V_s^n)+
\sum\limits_{t=1}^\infty\sum\limits_{j\in I^+}P(A_j^n \cap V_t^n)=1, \quad P  \in M.
\end{eqnarray}
Let us write inequality (\ref{33myk14}) in  the form
\begin{eqnarray}\label{3myk14}
\sum\limits_{s=1}^\infty\sum\limits_{i \in I^-} f_{is}^n P(A_i^n \cap V_s^n) +\sum\limits_{t=1}^\infty\sum\limits_{j \in I^+}f_{jt}^n P(A_j^n  \cap V_t^n) \leq 1, \quad P \in \bar M_n^U,
\end{eqnarray}
where $f_{is}^n=f_s^n, \ i=\overline{1,\infty},$ \  $ f_{jt}^n=f_t^n, \ j=\overline{1,\infty}.$

Due to completeness of the set of measures $M,$
for those $i \in I^-$ for which $d_{is}^n<0,  \ s=\overline{1, \infty}, $ and those $j \in I^+$ for which $d_{jt}^n>0,  \ t=\overline{1, \infty},$  the inequality (\ref{3myk14}) is as follows 
\begin{eqnarray}\label{3myk15}
 f_{is}^n \frac{d_{jt}^n}{-d_{is}^n +d_{jt}^n} + \frac{-d_{is}^n}{-d_{is}^n+d_{jt}^n} f_{jt}^n\leq 1, \quad d_{is}^n <0,  \ s=\overline{1, \infty}, \ 
\end{eqnarray} 
$$ d_{jt} ^n>0,  \ t=\overline{1, \infty}.$$
From  inequalities  (\ref{3myk15})  we obtain
\begin{eqnarray}\label{3myk16}
f_{jt}^n \leq 1+\frac{1- f_{is}^n}{-d_{is}^n}d_{jt}^n, \quad d_{is}^n <0, \   s=\overline{1, \infty}, \ d_{jt}^n >0,  \ t=\overline{1, \infty}.
\end{eqnarray} 
Two cases are possible: a)  $ f_{is}^n \leq 1, \ i \in I^-, \ s=\overline{1, \infty};  $ b) there exists $ i \in I^-$ such that $  f_{is}^n > 1, \ s=\overline{1, \infty}.$
Consider the case a).

Since inequalities  (\ref{3myk16}) are valid for every $\frac{1- f_{is}^n}{-d_{is}^n},$ as $ d_{is}^n <0,  \ s=\overline{1, \infty}, $ and $ f_{is}^n \leq 1, \  i \in I^-,  \ s=\overline{1, \infty},$  then if to denote
$$ \alpha_n =\inf_{\{is, \ d_{is}^n <0\}}\frac{1- f_{is}^n}{-d_{is}^n},$$
  we have  $ 0 \leq \alpha_n < \infty,$ and 
\begin{eqnarray}\label{3myk17}
f_{jt}^n \leq 1+\alpha_n d_{jt}^n,  \quad   d_{jt}^n >0, \quad  j \in I^+,  \quad  t=\overline{1, \infty}.
\end{eqnarray} 
From the definition of  $\alpha_n$ we obtain inequalities 
\begin{eqnarray*}
f_{is}^n \leq 1+\alpha_n d_{is}^n,  \quad   d_{is}^n <0, \quad  i \in I^-, \quad s=\overline{1, \infty}.
\end{eqnarray*} 
Now if $d_{is}^n=0$ for some $ i \in I^-, \ s=\overline{1, \infty},$ then in this case $f_{is}^n \leq 1.$ All these inequalities give 
\begin{eqnarray}\label{3myk18}
f_{is}^n \leq 1+\alpha_n d_{is}^n,  \quad   i \in I^-\cup I^+, \quad  s=\overline{1, \infty}.
\end{eqnarray} 
Consider the case b). From the  inequality (\ref{3myk16}) we obtain
\begin{eqnarray}\label{3myk19}
f_{jt}^n \leq 1-\frac{1- f_{is}^n}{d_{is}^n}d_{jt}^n, \ d_{is}^n <0, \ i  \in  I^-,  \   s=\overline{1, \infty}, 
\end{eqnarray}
\begin{eqnarray*}
\  d_{jt}^n >0, \ j \in I^+, \    t=\overline{1, \infty}. 
\end{eqnarray*}
The last inequality gives
\begin{eqnarray}\label{3myk20}
\frac{1- f_{is}^n}{d_{is}^n} \leq \min_{\{ jt, \ d_{jt}^n >0\}} \frac{1}{d_{jt}^n}< \infty, \ d_{is}^n <0, \ i \in I^-, \   s=\overline{1, \infty}.
\end{eqnarray}
Let us define $\alpha_n =\sup\limits_{\{ is, \ d_{is}^n<0\}}\frac{1- f_{is}^n}{d_{is}^n}< \infty.$
Then from (\ref{3myk19}) we obtain
\begin{eqnarray}\label{3myk21}
f_{jt}^n \leq 1- \alpha_n d_{jt}^n, \   d_{jt}^n >0, \ j \in I^+, \  t=\overline{1, \infty}.
\end{eqnarray}
From the definition of  $\alpha_n $ we obtain
\begin{eqnarray}\label{3myk22}
f_{is}^n \leq 1- \alpha_n d_{is}^n, \  d_{is}^n <0, \ i \in I^-, \  s=\overline{1, \infty}.
\end{eqnarray}
The inequalities (\ref{3myk21}), (\ref{3myk22}) give
\begin{eqnarray}\label{3myk23}
f_{is}^n \leq 1- \alpha_n d_{is}^n,  \quad  i \in I^-\cup I^+, \   s=\overline{1, \infty}.
\end{eqnarray}

   Multiplying on $\chi_{A_i^n \cap V_s^n}$ the inequalities (\ref{3myk23}) and summing over all $ i \in I^-\cup I^+$  and $ s=\overline{1, \infty}$  we obtain 
$$
\sum\limits_{s=1}^\infty\sum\limits_{i=1}^\infty f_{is}\chi_{A_i^n \cap V_s^n}=
\sum\limits_{s=1}^\infty\sum\limits_{i=1}^\infty f_{s}\chi_{A_i^n \cap V_s^n}=
\frac{\xi_n}{\sup\limits_{P \in M}E^p\xi_n} \leq $$
\begin{eqnarray}\label{33myk23}
1- \alpha_n \sum\limits_{s=1}^\infty\sum\limits_{i=1}^\infty d_{is}\chi_{A_i^n \cap V_s^n}=1- \alpha_n (m_n - m_{n-1}).
\end{eqnarray}

The Lemma  \ref{3myk10}  is proved.

\begin{thm}\label{3myk24} Suppose that conditions of the Lemma \ref{3myk10}
are valid. Then for every non negative supermartingale  $\{f_m, {\cal F}_m\}_{m=0}^\infty, $ satisfying conditions 
\begin{eqnarray}\label{3myk25}
\sup\limits_{P \in M}E^Pf_m < \infty, \quad  \frac{f_m}{f_{m-1}} \leq C_m< \infty, \quad m=\overline{1,\infty},
\end{eqnarray}
 optional decomposition is valid.
\end{thm}
{\bf Proof.} Consider random value $\xi_n=\frac{f_n}{f_{n-1}}.$ Due to Lemma \ref{3myk10}
$$ \frac{\xi_n}{\sup\limits_{P \in M}E^P\xi_n} \leq 1+\alpha_n(m_n-m_{n-1})=\xi_n^0.$$
It is evident that  $E^P\{\xi_n^0|{\cal F}_{n-1}\}=1, \ P \in M, \ n=\overline{1, \infty}.$ 
Since $\sup\limits_{P \in M}E^P\xi_n \leq 1,$ then 
\begin{eqnarray}\label{3myk26}
 \frac{f_n}{f_{n-1}} \leq \xi_n^0, \quad n=\overline{1, N}.
\end{eqnarray} 
The Theorem \ref{nick1} and inequalities (\ref{3myk26}) prove the Lemma \ref{3myk24}.

\section{ Application to Mathematical Finance.}

Due to Corollary \ref{mars16}, we can give the following definition of fair price of contingent claim $f_N$ relative to a convex set of equivalent measures $M.$
\begin{defin}\label{maras1}
Let $f_N, \ N< \infty, $ be a ${\cal F}_N$-measurable  integrable   relative to a convex set of equivalent measures  $M$  random value  such that for some  $0 \leq \alpha_0< \infty$ and $\xi_0 \in A_0$
 \begin{eqnarray}\label{maras2}
P(f_N - \alpha_0 E^P\{\xi_0|{\cal F}_N\} \leq 0)=1.
\end{eqnarray}
Denote $ G_{\alpha_0}=\{\alpha \in [0,  \alpha_0],  \ \exists \xi_{\alpha} \in A_0, \ P(f_N - \alpha E^P\{\xi_{\alpha}|{\cal F}_N\} \leq 0)=1\}.$
We call
\begin{eqnarray}\label{maras4}
f_0=\inf\limits_{\alpha \in G_{\alpha_0}}\alpha
\end{eqnarray}
  a fair price of  contingent claim  $f_N$ relative to a convex set of equivalent measures $M,$ if there exists $\zeta_0 \in A_0$  and a sequences $\alpha_n \in [0,\alpha_0],$  $\xi_{\alpha_n} \in A_0,$  satisfying conditions $\alpha_n \to f_0,$ 
$\xi_{\alpha_n} \to \zeta_0$ by probability, as $n \to \infty,$ and such that
\begin{eqnarray}\label{maras3}
P(f_N - \alpha_n E^P\{\xi_{\alpha_n}|{\cal F}_N\} \leq 0)=1, \quad n=\overline{1, \infty}.
\end{eqnarray}
\end{defin}
\begin{thm}\label{mars17}
Let the set $A_0$ be uniformly integrable one relative to every measure $P \in M.$ Suppose that  for  a nonnegative ${\cal F}_N$-measurable  integrable relative to every measure $P \in M$  contingent claim $f_N, \  N< \infty,$  there exist $\alpha_0< \infty$ and $\xi_0 \in A_0$ such that
\begin{eqnarray}\label{mars18}
P(f_N - \alpha_0 E^P\{\xi_0|{\cal F}_N\} \leq 0)=1,
\end{eqnarray}
then a fair price $f_0$ of contingent claim $f_N$ exists.
For $f_0$ the inequality 
\begin{eqnarray}\label{mars20}
 \sup\limits_{P \in M}E^Pf_N \leq f_0
\end{eqnarray}
is valid. If a supermartingale $\{f_m=\mathrm{ess}\sup\limits_{P \in M}E^P\{f_N|{\cal F}_m\},{\cal F}_m\}_{m=0}^\infty$ is  local regular one, then $f_0=\sup\limits_{P \in M}E^Pf_N.$ 
\end{thm}

{\bf Proof.} If $ f_0=\alpha_0,$ then Theorem \ref{mars17} is proved. Suppose that 
$ f_0<\alpha_0.$ Then there exists  a sequence $\alpha_n \to f_0,$ and $\xi_{\alpha_n} \in A_0, \  n \to \infty,$ such that
\begin{eqnarray}\label{mars21}
 P(f_N - \alpha_n E^P\{\xi_{\alpha_n}|{\cal F}_N\} \leq 0)=1, \quad P \in M.
\end{eqnarray}
 Due to uniform integrability $ A_0$ we obtain
\begin{eqnarray}\label{mars22}
1=\lim\limits_{n \to \infty}\int\limits_{\Omega}\xi_{\alpha_{n}}dP=\int\limits_{\Omega} \zeta_0 dP, \quad P \in M.
\end{eqnarray}
 Using again uniform integrability and going to the limit in   (\ref{mars21}) we obtain
\begin{eqnarray}\label{mars23}
 P(f_N - f_0 E^P\{ \zeta_0|{\cal F}_N\} \leq 0)=1, \quad P \in M.
\end{eqnarray}
From the inequality $ f_N - f_0 E^P\{ \zeta_0|{\cal F}_N\} \leq 0$ it follows
 inequality (\ref{mars20}).
If $f_m=\mathrm{ess}\sup\limits_{P \in M}E^P\{f_N|{\cal F}_m\}, \  m=\overline{0,N},$ is a local regular supermartingale, then
\begin{eqnarray}\label{bars1}
f_m=M_m- g_m, \quad m=\overline{0,N}, \quad g_0=0,
\end{eqnarray}
where a martingale $M_m, \ m=\overline{0,N},$ is a nonnegative and $E^P M_m=\sup\limits_{P \in M}E^P f_N.$ Introduce  into consideration a random value $\xi_0=\frac{M_N}{\hat f_0}, \hat f_0= \sup\limits_{P \in M}E^Pf_N. $ Then $\xi_0$ belongs to the set $A_0$ and 
\begin{eqnarray}\label{bars2}
P(f_N - \hat f_0  E^P\{ \xi_0|{\cal F}_N\} \leq 0)=1.	
\end{eqnarray}
From this it follows that
 $f_0=\sup\limits_{P \in M}E^Pf_N.$ 

Let us prove that $f_0$ is a fair price for some evolution of risk and non risk assets.
Suppose that evolution of risk asset is given by the law $S_m=f_0M^P\{\zeta_0|{\cal F}_m\}, \ m=\overline{0,N}, $ and evolution of non risk asset is given by the formula  $B_m=1,  \ m=\overline{0,N}.$

As proved above,  for $f_0=\inf\limits_{\alpha \in G_{\alpha_0}}\alpha $
there exists $ \zeta_0 \in A_0$  such that the inequality 
\begin{eqnarray*}\label{ghon2}
f_N - f_0 E^P\{\zeta_0|{\cal F}_N\} \leq 0
\end{eqnarray*}
is valid. Let us put 
$$f_m^0=f_0 E^P\{\zeta_0|{\cal F}_m\},  \quad P \in M,$$
$$\bar f_m=
\left\{\begin{array}{l l} 0, & m<N, \\
f_N -f_0 E^P\{\zeta_0|{\cal F}_m\}, & m = N.  
 \end{array} \right. $$
It is evident that $\bar f_{m-1} -\bar f_m \geq 0, \  m=\overline{0, N}.$
Therefore, the supermartingale
$$f_m^0+ \bar f_m=
\left\{\begin{array}{l l} f_0 E^P\{\zeta_0|{\cal F}_m\}, & m<N, \\
f_N , & m= N,  
\end{array} \right. $$
is a local regular one.
It is evident that
$$f_m^0+ \bar f_m=M_m - g_m, \quad  m=\overline{0, N},$$ 
where $$ M_m=f_0 E^P\{\zeta_0|{\cal F}_m\},  \quad  m=\overline{0, N},$$
$$ g_m = 0, \quad  \ m=\overline{0, N-1},$$ 
$$ g_ N =f_0 E^P\{\zeta_0|{\cal F}_N\} - f_N.$$
  For martingale  $\{M_m\}_{m=0}^N$ the representation
$$ M_m=f_0+\sum\limits_{i=1}^m H_i \Delta  S_i, \quad m=\overline{0, N},$$
 is valid, where $H_i=1, \  i=\overline{1, N}.$  
Let us consider a trading strategy  $\pi=\{\bar H_m^0, \bar H_m\}_{m=0}^N,$ where
$$\bar H_0^0=f_0, \quad \bar H_m^0=M_m - H_m S_m, \quad m=\overline{1,N}, \quad  \bar H_0=0, \quad \bar H_m=H_m, \quad m=\overline{1,N}.$$
It is evident  that  $\bar H_m^0,  \bar H_m$ are  ${\cal F}_{m-1}$ measurable and the trading strategy  $\pi$ satisfy  self-financed condition
$$ \Delta \bar H_m^0 +\Delta  \bar H_m S_{m-1}=0.$$
Moreover, a capital corresponding to the  self-financed trading strategy  $\pi$ is given by the formula
$$X_m^{\pi}=\bar H_m^0+  \bar H_m  S_{m} = M_m.$$
Herefrom,  $X_0^{\pi}=f_0.$
Further,
$$ X_N^{\pi}=f_N+g_N \geq f_N.$$

The last proves the Theorem \ref{mars17}.
 From (\ref{mars23}) and Corollary \ref{mars16} the Theorem \ref{mars26}  follows.

\begin{thm}\label{mars26} 
Suppose that the set $A_0$ contains  only $ 1 \leq  k < \infty $ linear independent elements $\xi_1, \ldots \xi_k.$ If there exist $\xi_0 \in T$ and $\alpha_0\geq 0$ such that
\begin{eqnarray}\label{00mars27}
P(f_N - \alpha_0E^P\{ \xi_0 | {\cal F}_N \}\leq 0)=1, \quad P \in M,
\end{eqnarray}
where
\begin{eqnarray}\label{mars28}
T=\{\xi \geq 0, \  \xi=\sum\limits_{i=1}^k\alpha_i\xi_i, \ \alpha_i \geq 0, \ i=\overline{1,k}, \ \sum\limits_{i=1}^k\alpha_i=1\},
\end{eqnarray}
then a fair price $f_0$ of contingent claim $f_N \geq 0$ exists, where $f_N$ is ${\cal F}_N$ measurable and integrable relative to every measure $P \in M,$ $N< \infty. $
\end{thm}
{\bf Proof.} The proof is evident, as the set $T$ is uniformly integrable relative to every measure from $M.$

\begin{cor}\label{rian1} On a measurable space $\{\Omega, {\cal F}\}$ with filtration ${\cal F}_m$ on it,  let  \\ $\{f_m, {\cal F}_m\}_{m=0}^N$ be a non negative  local regular supermartingale   relative to a convex set of equivalent measures $M.$   If the set $A_0$ is uniformly integrable relative to every measure $P \in M,$ then the fair price of contingent claim $f_N$ exists.
\end{cor}
{\bf Proof.} From optional decomposition  we have $f_m=M_m - g_m, \ m=\overline{0,N}.$ Therefore, $P(f_N - \alpha_0 \xi_0 \leq 0)=1,$ where $\alpha_0=E^PM_N, \  P \in M, \xi_0=\frac{M_N}{E^PM_N}. $ From the last it follows that conditions of Theorem \ref{mars26} are satisfied. The Corollary \ref{rian1} is proved.

On a probability space $\{\Omega, {\cal F}, P\}$ let us   consider  an evolution of  one risk asset    given by the law $ \{ S_m \}_{m=0}^N,$  where $S_m$ is a random  value taking values in $R_+^1.$  Suppose that
 ${\cal F}_m$ is a filtration on  $\{\Omega, {\cal F}, P\}.$ 
 We assume that non risk asset evolve by the law $B_m^0=1,  \ m=\overline{1,N}. $
Denote by $M^e(S)$ the set of all martingale measures being equivalent to the measure $P.$ 
 We assume that the set  $M^e(S)$ of such  martingale measures is not empty and effective market is non complete, see, for example, \cite{DMW90}, \cite{K81}, \cite{Schacher1}, \cite{HK79}.
So, we have that 
\begin{eqnarray}\label{ma29}
E^Q\{S_m|{\cal F}_{m-1}\}=S_{m-1}, \quad m=\overline{1, N}, \quad Q \in M^e(S).
\end{eqnarray}

The next Theorem justify the Definition \ref{maras1}.

\begin{thm}\label{hon1} Let a contingent claim $f_N $  be  a 
${\cal F}_N $-measurable   integrable random value with respect to every measure from $M^e(S)$ and conditions of the Theorem  \ref{mars26} are satisfied with $\xi_i=\frac{S_i}{S_0}, \ i=\overline{0,N}.$ Then  there exists self-financed trade strategy $\pi$  the capital  evolution $\{X_m^{\pi}\}_{m=0}^N$ of which  is  a martingale relative to every measure from $M^e(S)$  satisfying conditions  $ X_0^{\pi}=f_0, \  X_N^{\pi} \geq f_N,$
where  $f_0$ is a fair price of contingent claim $f_N. $ 
\end{thm}
{\bf Proof.} Due to Theorems \ref{mars17}, \ref{mars26}, for $f_0=\inf\limits_{\alpha \in G_{\alpha_0}}\alpha $
there exists $ \zeta_0 \in A_0$  such that the inequality 
\begin{eqnarray}\label{ghon2}
f_N - f_0 E^P\{\zeta_0|{\cal F}_N\} \leq 0
\end{eqnarray}
is valid. Let us put 
$$f_m^0=f_0 E^P\{\zeta_0|{\cal F}_m\},  \quad P \in M^e(S),$$
$$\bar f_m=
\left\{\begin{array}{l l} 0, & m<N, \\
f_N -f_0 E^P\{\zeta_0|{\cal F}_m\}, & m = N.  
 \end{array} \right. $$
It is evident that $\bar f_{m-1} -\bar f_m \geq 0, \  m=\overline{0, N}.$
Therefore, the supermartingale
$$f_m^0+ \bar f_m=
\left\{\begin{array}{l l} f_0 E^P\{\zeta_0|{\cal F}_m\}, & m<N, \\
f_N , & m= N,  
\end{array} \right. $$
is a local regular one.
It is evident that
$$f_m^0+ \bar f_m=M_m - g_m, \quad  m=\overline{0, N},$$ 
where $$ M_m=f_0 E^P\{\zeta_0|{\cal F}_m\},  \quad  m=\overline{0, N},$$
$$ g_m = 0, \quad  \ m=\overline{0, N-1},$$ 
$$ g_ N =f_0 E^P\{\zeta_0|{\cal F}_N\} - f_N.$$
Due to Theorem \ref{t8},  for martingale  $\{M_m\}_{m=0}^N$ the representation
$$ M_m=f_0+\sum\limits_{i=1}^m H_i \Delta  S_i, \quad m=\overline{0, N},$$
 is valid. 
Let us consider a trading strategy  $\pi=\{\bar H_m^0, \bar H_m\}_{m=0}^N,$ where
$$\bar H_0^0=f_0, \quad \bar H_m^0=M_m -  H_m   S_m, \quad m=\overline{1,N}, \quad  \bar H_0=0, \quad \bar H_m=H_m, \quad m=\overline{1,N}.$$
It is evident  that  $\bar H_m^0,  \bar H_m$ are  ${\cal F}_{m-1}$ measurable and the trading strategy  $\pi$ satisfy  self-financed condition
$$ \Delta \bar H_m^0 +\Delta  \bar H_m  S_{m-1}=0.$$
Moreover, a capital corresponding to the  self-financed trading strategy  $\pi$ is given by the formula
$$X_m^{\pi}=\bar H_m^0+  \bar H_m  S_{m} = M_m.$$
Herefrom,  $X_0^{\pi}=f_0.$
Further,
$$ X_N^{\pi}=f_N+g_N.$$
Therefore $X_N^{\pi} \geq f_N.$ Theorem \ref{hon1}  is proved.

In the next theorem we assume that evolutions of risk and non risk assets generate  incomplete  market   \cite{DMW90}, \cite{K81}, \cite{Schacher1},  \cite{HK79}, that is, the set of martingale measures contains more that one element.

\begin{thm}\label{mars30}
Let  an evolution $ \{ S_m \}_{m=0}^N$ of risk asset  satisfy conditions $P(D_m^1 \leq S_m \leq D_m^2)=1, \ D_{m-1}^1 \geq  D_m^1>0, \  D_{m-1}^2 \leq D_{m}^2< \infty, \ m=\overline{1,N},$ and let non risk asset evolution be deterministic one given  by the law $\{B_m\}_{m=0}^N, \ B_m=1, \ m=\overline{0, N}.$ The fair price of standard European call option  with payment function $f_N=(S_N - K)^+$
is given by the formula
\begin{eqnarray}\label{mars31}
f_0=\left\{\begin{array}{l l} S_0(1-\frac{K}{D_N^2}), & K \leq D_N^2, \\
 0, & K >D_N^2. \end{array}\right.
\end{eqnarray}

The fair price of standard European put option  with payment function $f_N=(K-S_N)^+$
is given by the formula
\begin{eqnarray}\label{mmars31}
f_0=\left\{\begin{array}{l l} K- D_N^1, & K \geq D_N^1, \\
 0, & K < D_N^1. \end{array}\right.
\end{eqnarray}
\end{thm}
{\bf Proof.} In the Theorem \ref{mars30} conditions the set of equations $E^P\zeta=1, \ \zeta \geq 0, $ has  solutions $\zeta_i=\frac{S_i}{S_0}, \ i=\overline{0,N}.$ It is evident that  $\alpha_0=S_0$ and $\zeta_N=\frac{S_N}{S_0},$ since
$$ \frac{(S_N - K)^+}{B_N} - \alpha_0 \frac{S_N}{S_0} \leq 0, \quad \omega \in \Omega.$$ 
Let us prove the needed formula. Consider the inequality
\begin{eqnarray}\label{mars33}
  (S_N - K) - \alpha \sum\limits_{i=0}^N\gamma_i\frac{S_i}{S_0} \leq 0, \quad  \gamma \in V_0,
\end{eqnarray}
where  $V_0=\{\gamma=\{\gamma_i\}_{i=0}^N, \ \gamma_i \geq 0, \ \sum\limits_{i=0}^N\gamma_i=1\}.$
Or,
\begin{eqnarray}\label{mars34}
 S_N\left ( 1 - \frac{ \alpha\gamma_N}{S_0}\right) - K - \alpha \sum\limits_{i=0}^{N-1}\gamma_i\frac{S_i}{S_0} \leq 0.
\end{eqnarray}
Suppose that  $\alpha$ satisfies inequality
\begin{eqnarray}\label{mars35}
 1 -  \frac{ \alpha}{S_0}>0.
\end{eqnarray}
If $\alpha $ satisfies additionally the equality
\begin{eqnarray}\label{mars36}
 D_N^2\left( 1 - \frac{ \alpha\gamma_N}{S_0}\right) - K - \alpha \sum\limits_{i=0}^{N-1}\gamma_i\frac{D_i^1}{S_0}= 0,
\end{eqnarray}
  then for all $\omega \in \Omega$ 
(\ref{mars34}) is valid.  From (\ref{mars36}) we obtain for $\alpha$
\begin{eqnarray}\label{mars37}
\alpha=\frac{S_0(D_N^2 - K)}{(D_N^2\gamma_N+ \sum\limits_{i=0}^{N-1}\gamma_i D_i^1)}.
\end{eqnarray}
If $D_N^2 -K >0,$ then
\begin{eqnarray}\label{mars38}
\inf\limits_ {\gamma \in V_0}\frac{S_0(D_N^2 - K)}{(D_N^2\gamma_N+ \sum\limits_{i=0}^{N-1}\gamma_i D_i^1)}=\frac{S_0(D_N^2 - K)}{D_N^2},
\end{eqnarray}
since $ D_N^2 \geq D_i^1.$
From here we obtain
\begin{eqnarray}\label{mars39}
f_0=S_0\left(1 - \frac{ K}{D_N^2}\right).
\end{eqnarray}
It is evident that $\alpha = f_0$ satisfies inequality (\ref{mars35}).

If  $D_N^2 - K \leq 0,$ then $S_N - K \leq 0$  and from (\ref{mars33}) we can put $\alpha=0.$ Then, the formula (\ref{mars34}) is valid for all $\omega \in \Omega.$

Let us prove the formula (\ref{mmars31}) for standard European put option.
If $S_N \leq K$ it is evident that  $\alpha_0=K, $ and $\zeta_0=1,$ since
$$ (K -S_N) - \alpha_0  \leq 0, \quad \omega \in \Omega.$$ 
Let us prove the needed formula. 
 Consider the  inequality
\begin{eqnarray}\label{mmars33}
  (K- S_N)^+ - \alpha \sum\limits_{i=0}^N\gamma_i\frac{S_i}{S_0}  \leq 0,  \quad  \gamma \in V_0.
\end{eqnarray}
Or, for $S_N \leq K$
\begin{eqnarray}\label{mmars34}
 - S_N\left( 1 + \frac{ \alpha \gamma_N}{S_0}\right) + K -\alpha \sum\limits_{i=0}^{N-1}\gamma_i\frac{S_i}{S_0}\leq 0.
\end{eqnarray}
If  $\alpha$ is a solution of the equality
\begin{eqnarray}\label{mmars36}
 - D_N^1\left( 1 + \frac{ \alpha \gamma_N}{S_0}\right) + K - \alpha \sum\limits_{i=0}^{N-1}\gamma_i\frac{D_i^1}{S_0}  = 0,
\end{eqnarray}
 then for all $\omega \in \Omega$ 
(\ref{mmars34}) is valid.  From (\ref{mmars36}) we obtain for $\alpha$
\begin{eqnarray}\label{mmars37}
\alpha=\frac{S_0( K - D_N^1)}{\sum\limits_{i=0}^{N}\gamma_i D_i^1 }.
\end{eqnarray}
Therefore,
\begin{eqnarray}\label{mmars38}
\inf\limits_ {\gamma \in V_0}\frac{S_0(K- D_N^1)}{\sum\limits_{i=0}^{N}\gamma_i D_i^1}= K - D_N^1,
\end{eqnarray}
since $ D_i^1\leq S_0, \ i=\overline{1,N}, \ D_0^1=S_0.$
From here we obtain
\begin{eqnarray}\label{mmars39}
f_0= K - D_N^1.
\end{eqnarray}
If  $D_N^1 - K > 0,$ then $S_N - K > 0$  and from (\ref{mmars33}) we can put $\alpha=0.$ Then, (\ref{mmars34}) is valid for all $\omega \in \Omega.$
The Theorem  \ref{mars30} is proved.

\section{Some auxiliary results.}
On a measurable space $\{\Omega, {\cal F}\}$ with filtration ${\cal F}_n$ on it,  let us consider  a convex set of equivalent measures $M.$
Suppose that  $\xi_1, \ldots, \xi_d$ is a set of random values belonging to the set $A_0.$ 
Introduce $d$ martingales relative to a set of measures $M$ $\{S_n^i, {\cal F}_n\}_{n=0}^{\infty}, \  i=\overline{1,d},$    where  $S_n^i=E^P\{\xi_i|{\cal F}_n\}, \  i=\overline{1,d}, \ P \in M. $    Denote by    $M^{e}(S)$   a set of all  equivalent to  a measure  $P \in M$  martingale measures,   that is,  $Q \in M^{e}(S)$ if
$$ E^Q\{S_n|{\cal F}_{n-1}\}=S_{n-1},\quad E^Q|S_n|< \infty,  \quad Q \in M^{e}(S), \quad n=\overline{1, \infty}. $$ 
It is evident that  $ M \subseteq M^{e}(S) $  and  $M^{e}(S)$ is a convex set.
Denote by   $ P_0$  a certain fixed measure from $M^{e}(S)$  and let  $L^0(R^d)$   be a set of finite valued random values on  a probability space $\{ \Omega, {\cal F}, P_0\},$ taking values in $R^d.$

 Let    $H^0$ be  a set of finite valued predictable processes $H=\{H_n\}_{n=1}^N, $ where $H_n=\{H_n^i\}_{i=1}^d$ takes values in $R^d$  and  $ H_n$ is ${\cal F}_{n-1}$-measurable.
Introduce into consideration a set of random values  
\begin{eqnarray}\label{s5}
K_N^1=\{ \xi \in L^0(R^1), \  \xi= \sum\limits_{k=1}^N   \langle H_k,\Delta S_k \rangle, , \ H  \in H^0\}, \quad  N< \infty,
\end{eqnarray}

$$\Delta S_k = S_k - S_{k-1}, \quad \langle H_k, \Delta S_k \rangle = \sum\limits_{s=1}^dH_k^s(S_k^s-S_{k-1}^s).$$ 

\begin{lemma}\label{q7}
The set of random values  $K_N^1$  is  a closed subset in the set of finite valued random values  $L^0(R^1)$  relative to convergence by measure $P \in M.$
\end{lemma}
 The proof of the Lemma \ref{q7} see, for example, \cite{DMW90}.

Introduce into consideration a subset
 $$V^0 =\{H \in H^0,\ ||H_n||< \infty,\ n=\overline{1, N}\} $$ 
of the set    $H^0,$ where $||H_n||=\sup\limits_{\omega \in \Omega}\sum\limits_{i=1}^d|H_n^i|.$ 
Let $K_N$ be a subset of the set $K_N^1$
$$ K_N=\{\xi \in L^0(R^1), \  \xi=\sum\limits_{k=1}^N   \langle H_k,\Delta S_k \rangle, \ H \in V^0 \}.$$

Denote also a set 
$$C =\{k-f,\ k \in  K_N, \ f \in L_+^{\infty}(\Omega, {\cal F}, P_0)\},$$
where  $L_+^{\infty}(\Omega, {\cal F}, P_0\}$  is a set of bounded nonnegative random values. Let   $\bar C$ be   a closure of  $C$ in $L^{1}(\Omega, {\cal F}, P_0)$ metrics.
\begin{lemma}\label{l8} If  $\zeta \in \bar C$  and such that  $E^{P_0}\zeta=0,$ then for   $\zeta$ the representation  
$$\zeta=\sum\limits_{k=1}^N \langle H_k,\Delta S_k \rangle $$
is valid for a certain finite valued predictable process
  $H=\{H_n\}_{n=1}^N.$
\end{lemma}
{\bf Proof.} If $\zeta \in K_N,$  then Lemma \ref{l8} is proved. Suppose that   $\zeta \in \bar C,$  then there exists a sequence  $k_n - f_n, \ k_n \in K_N,\ f_n \in  L_+^{\infty}(\Omega, {\cal F}, P_0)$ such that 
$||k_n - f_n - \zeta ||_{P_0} \to 0,\ n \to \infty,$ where $||g||_{P_0}=E^{P_0}|g|.$ Since 
$|E^{P_0}(k_n - f_n - \zeta)| \leq ||k_n - f_n - \zeta ||_{P_0},$ we have $E^{P_0}f_n  \leq ||k_n - f_n - \zeta ||_{P_0}.$
From here we obtain    $||k_n - \zeta ||_{P_0} \leq 2 ||k_n - f_n - \zeta ||_{P_0}.$ Therefore, $k_n \to \zeta$
by measure $P_0.$  On the basis of the Lemma  \ref{q7}, a set 
$$ K_N^1=\{\xi \in L^0(R^1), \  \xi=\sum\limits_{k=1}^N \langle H_k,\Delta S_k \rangle, \ H \in H^0 \}, \  \langle H_k,\Delta S_k \rangle=\sum\limits_{i=1}^dH_k^i(S_k^i - S_{k-1}^i)$$
is  a closed subset of $L^0(R^1)$  relative to convergence by measure  $P_0.$ 
From this fact, we obtain the proof of Lemma  \ref{l8}, since there exists finite valued predictable process  $H \in H^0$ such that for   $\zeta $ the representation 
$$\zeta=\sum\limits_{k=1}^N\langle H_k,\Delta S_k \rangle $$
is valid.

\begin{thm}\label{t7} Let  $E^Q|\zeta| < \infty, \ Q \in  M^{e}(S).$ If for every    $\ Q \in  M^{e}(S), \ E^Q\zeta = 0,$ then there exists  finite valued predictable process  $H$ such that for  $ \zeta$ the representation  
\begin{eqnarray}\label{m1}
 \zeta=\sum\limits_{k=1}^N \langle H_k,\Delta S_k \rangle
\end{eqnarray}
is valid.
\end{thm}
{\bf Proof.}  If  $\zeta  \in \bar C,$ then  (\ref{m1}) follows from Lemma  \ref{l8}. So, let  $\zeta$  does not belong to $\bar C.$  As in Lemma  \ref{l8}, $\bar C$  is a closure of  $C$ in  $L^{1}(\Omega, {\cal F}, P_0)$  metrics for the  fixed measure  $P_0.$ The set  $\bar C$ is a closed convex set in $L^{1}(\Omega, {\cal F}, P_0).$   Consider the  other convex closed set that consists from one element   $\zeta. $ Due to  Han -- Banach  Theorem, there exists a linear continuous functional $l_1,$ which belongs to  $L^{\infty}(\Omega, {\cal F}, P_0),$  and real numbers  $\alpha > \beta$ such that 
\begin{eqnarray}\label{m2}
l_1(\xi)=\int\limits_{\Omega}\xi(\omega) q(\omega)dP_0, \quad q(\omega) \in  L^{\infty}(\Omega, {\cal F}, P_0),
\end{eqnarray}
and inequalities  $l_1(\zeta) > \alpha,$ $l_1(\xi) \leq \beta, \ \xi \in  \bar C,$  are valid.  Since   $\bar C$  is a convex cone we can put $ \beta=0.$  From condition  $l_1(\xi) \leq 0, \ \xi \in  \bar C$ we have $l_1(\xi)=0, \  \xi \in K_N^1 \cap L^{1}(\Omega, {\cal F}, P_0).$ From  (\ref{m2}) and inclusions $ \bar C \supset C \supset - L^{\infty}(\Omega, {\cal F}, P_0)$ we have  $ q(\omega) \geq 0.$  Introduce  a measure
$$Q^*(A)= \int\limits_{A} q(\omega)dP_0\left[\int\limits_{\Omega} q(\omega)dP_0\right]^{-1}.$$
Then, we have 
\begin{eqnarray}\label{m3}
  \int\limits_{\Omega} \xi(\omega)dQ^*=0,\quad \xi \in K_N^1 \cap L^{1}(\Omega, {\cal F}, P_0).
\end{eqnarray}
Let us choose  $ \xi=\chi_{A}(\omega)(S_i^j -S_{i-1}^j ), \ A \in {\cal F}_{i-1},$ where  $\chi_{A}(\omega)$ is an indicator of a set   $A.$  We obtain 
$$\int\limits_{A}(S_i^j -S_{i-1}^j )dQ^*=0, \quad  A \in {\cal F}_{i-1}.$$
So, $Q^*$ is a martingale measure that belongs to the set  $M^a(S),$ which is a set of absolutely continuous martingale measures.  Let us choose   $Q \in  M^e(S)$ and consider a measure  $Q_1=(1- \gamma)Q+\gamma Q^* , \ 0 < \gamma < 1.$ A measure   $ Q_1 \in  M^e(S) $ and, moreover, $E^{Q_1}\zeta=\gamma E^{Q^*}\zeta>0.$  We come to the contradiction  with  conditions of Theorem  \ref{t7}, since for  $Q \in  M^e(S), \ E^Q\zeta=0.$ So, $\zeta \in \bar C,$  and in accordance with the Lemma  \ref{l8}, for  $\zeta$ the  declared  representation in Theorem \ref{t7}  is valid.

\begin{thm}\label{t8}
 For every martingale  $\{M_n,{\cal F}_n\}_{n=0}^{\infty}$ relative to the set of measures  $ M^e(S),$ there exists a predictable random process
$H$ such that for   $M_n,\ n=\overline{0,\infty},$ the representation 
\begin{eqnarray}\label{m4}
M_n=M_0+\sum\limits_{i=1}^n\langle H_i,\Delta S_i \rangle, \quad  n=\overline{1,\infty},
\end{eqnarray}
is valid.
\end{thm}
{\bf Proof.}  For fixed natural  $N \geq 1,$ let us consider random value $M_N - M_0=\zeta.$  Since $$E^Q|\zeta|<\infty, \quad E^Q\zeta=0, \quad Q \in  M^e(S),$$  then 
  $\zeta$ satisfies conditions of Theorem \ref{t7}  and, therefore,  belongs to  $\bar C,$ so, there exists a sequence   
$k_n=\sum\limits_{i=1}^N \langle H_i^n, \Delta S_i \rangle \in K_N $ such that 
$$\int\limits_{\Omega}|k_n - \zeta|dP_0 \to 0, \quad n \to \infty.$$
From here, we obtain 
$$\int\limits_{\Omega}|E^{P_0}\{(k_n - \zeta)|{\cal F}_{m}\}|dP_0 \leq \int\limits_{\Omega}|k_n - \zeta|dP_0 \to 0,\quad n \to \infty. $$
But  $E^{P_0}\{k_n |{\cal F}_{m}\}=\sum\limits_{i=1}^m \langle H_i^n, \Delta S_i \rangle.$
Hence,  we obtain that   as  $\sum\limits_{i=1}^m\langle H_i^n, \Delta S_i\rangle $ and  $\sum\limits_{i=1}^N \langle H_i^n, \Delta S_i \rangle $
converges by measure  $P_0$ to  $E^{P_0}\{\zeta|{\cal F}_{m}\}$ and  $\zeta,$ correspondingly.
There exists a subsequence  $n_k$ such that $H^{n_k}$  converges everywhere to predictable  process   $H$.
From here, we have  $\zeta = \sum\limits_{i=1}^N\langle H_i, \Delta S_i\rangle $ and  $E^{P_0}\{\zeta|{\cal F}_{m}\}=\sum\limits_{i=1}^m\langle H_i, \Delta S_i\rangle.$  It proves that for all   $m< N$
$$ M_m=M_0+\sum\limits_{i=1}^m\langle H_i, \Delta S_i \rangle.$$
Theorem  \ref{t8}  is proved.

\section{Conclusions.}

In the paper, we generalize Doob  decomposition
  for supermartingales relative to one measure  onto the case of supermartingales relative to a convex set of equivalent measures.  For supermartingales relative to one measure for  continuous time Doob's result was generalized in papers 
\cite{Meyer1} \cite{Meyer2}.

 Section 2  contains   the auxiliary statements giving sufficient conditions 
of the existence of maximal element in a maximal chain, of the existence of nonzero non-decreasing process such that the sum of a supermartingale and this process is again a supermartingale relative to a convex set of equivalent measures needed for the main Theorems. In  Theorem \ref{ctt5} we give sufficient conditions of the existence of the optional  Doob  decomposition for the special case as the set  of measures  is generated  by finite set of equivalent  measures  with bounded as below and above the Radon - Nicodym derivatives. After that, we introduce the notion of a regular supermartingale. Theorem \ref{ct4} describes   regular supermartingales. In  Theorem \ref{reww1} we give the necessary and sufficient conditions of regularity of supermartingales.
  After that we introduce a notion of local regular supermartingale.
At last, we prove  Theorem \ref{hf1} asserting that if the optional decomposition for a supermartingale is valid, then it is local regular one. 
Essentially, Theorem \ref{hf1} and \ref{mars1} give the necessary and sufficient conditions of local regularity of supermartingale.

 In section 3 we prove auxiliary statements nedeed for the description of local regular supermartingales.
 The notion of a local regular supermartingale relative to a convex set of equivalent measures  is equivalent to the existence of non negative  adapted process such that the equalities (\ref{mars2}) are valid.  Since the existence of optional decomposition  for supermartingale  and existence of  adapted non negative process entering  (\ref{mars2})  are equivalent ones,  then it would seem to obtain  new information  from the set of equation (\ref{mars2}) is impossible. As it was found,  this new formulation are proved to be  fruitful, since it turned out to describe  the structure of  all local regular supermartingales relative to a convex set of equivalent measures.
For this purpose we investigate the structure of supermartingales of special types
relative to a convex set of equivalent measures, generated by a certain finite set of equivalent measures. The main result of this investigation   is  the Lemma \ref{q5}, which allowed us to prove Lemma \ref{1q5}, stating sufficient conditions of existence of a martingale on a measurable space with respect to a convex set of equivalent measures generated by finite set of equivalent measures. The existence of non trivial random value  satisfying conditions (\ref{r7}) is sufficient condition for the existence of non trivial martingale with respect to a convex set of equivalent measures, generated by finite set of equivalent measures.
Theorem \ref{mars12} describes all local regular non negative supermartingales  of special type (\ref{marsss13})  relative to constructed above  set of equivalent measures.

In the Theorem \ref{fmars5} we give sufficient conditions  of the existence  of local regular martingale relative to arbitrary set of equivalent measures and arbitrary filtration. If time interval is finite  these conditions are also necessary. After that, we present in Theorem \ref{mmars1} important  construction of local regular supermartingales which  we sum up in Corollary \ref{fdr1}. Theorem \ref{mmars9} proves that every non negative uniformly integrable supermartingale  belongs to described class (\ref{mmars88}) of local regular supermartingales. 

Section 4 contains  the Theorem  \ref{nick1} giving a variant of the necessary and sufficient conditions of local regularity of non negative  supermartingale relative to a convex set of equivalent measures.   In subsection 1  the Definition \ref{1myk8} determine a class  of complete set  of equivalent measures.  The Lemma \ref{1myk10} guarantee   a bound  (\ref{1myk11})  for all non negative random values  allowing us  to prove  the Theorem \ref{1myk19}, stating that for every non negative supermartingale optional decomposition is valid.
In subsection 2 we extend the results of subsection 1 onto the case as a space of elementary events is countable.  At last,  subsection 3 contains the generalization of the result obtained in subsection 2 onto the case of arbitrary space of elementary events. We prove that for  every non negative supermartingale optional decomposition is valid.

 Corollary \ref{mars16}  of the Section 5 contains important construction of the local regular supermartingales playing important role in definition of fair price of contingent claim relative to a convex set of equivalent measures.
The Definition \ref{maras1} is fundamental for evaluation of risk in incomplete markets. 
Theorem \ref{mars17} gives sufficient conditions of the existence of fair price  of contingent claim relative to a convex set of equivalent measures. It also gives sufficient conditions  when defined fair price coincides with classical value. In the Theorem \ref{mars26} simple conditions of the existence  of fair price  of contingent claim  are given. In  Theorem \ref{hon1}  we  prove  the existence  of self-financed trading  strategy confirming a Definition \ref{maras1} of fair price  as parity between long and short positions in contracts. 
As application of the result obtained we prove Theorem \ref{mars30}, where the formulas for standard European call and put options in incomplete market we present. 
Section 6 contains auxiliary results needed for previous sections.

\vskip 5mm

\end{document}